\documentclass[preprint,12pt]{elsarticle}

\usepackage{amssymb}

\usepackage{graphicx}
\usepackage{amsmath}
\usepackage{amsthm}
\usepackage{subfigure}
\usepackage{epstopdf}
\usepackage{epsfig}

\begin{document}

\begin{frontmatter}



\title{ Multi-UAV-based Optimal Crop-dusting of Anomalously Diffusing Infestation of Crops }
\author[shu,ucm]{Jianxiong Cao}
\ead{caojianxiong2007@126.com}
 \author[ucm]{YangQuan Chen \corref{cor1}}
 \ead{ychen53@ucmerced.edu}
 \author[shu]{Changpin Li }
 \ead{lcp@shu.edu.cn}
 \address[shu]{Department of Mathematics, Shanghai University, Shanghai 200444, China}
 \address[ucm]{ School of Engineering, University of California, Merced, 5200 North Lake Road, Merced, CA 95343, USA}
 \cortext[cor1]{Corresponding author}

\begin{abstract}
This paper presents a UAV-based optimal crop-dusting method to control anomalously diffusing infestation of crops. Two anomalous diffusion models are considered, which are, respectively, time-fractional order diffusion equation and space-fractional order diffusion equation. Our problem formulation is motivated by real-time pest management by using networked unmanned cropdusters where the pest spreading is modeled as a fractional diffusion equation. We attempt to solve the optimal dynamic location of actuators by using Centroidal Voronoi Tessellations. A new simulation platform (FO-DiffMAS-2D) for measurement scheduling and controls in fractional order distributed parameter systems is also introduced in this paper. Simulation results are presented to show the effectiveness of the proposed method as well as the role of fractional order in the overall control performance.
\end{abstract}

\end{frontmatter}



\section{Introduction}
  It is well known that diffusion model is one of the most commonly used mathematical approaches for the description of transport dynamics, and the normal diffusion equation is used to govern the corresponding process. However, it fails to describe some transport dynamics in various complex systems. In the past few  decades, fractional calculus has attracted an increasing interests of researchers, and by using fractional calculus, anomalous diffusion models are developed to describe transport process in complex dynamic system \cite{metzler2000random}. Due to the nonlocal and hereditary properties of fractional operators, fractional diffusion equations are used to simulate those anomalous diffusion transport dynamics in complex systems. In this paper, we study the complex transport dynamic behavior of pest infestation of crops. One of the main problems is to locate the diffusion source, give optimal dynamic actuator location according to the measured data from sensors, and then take action to the anomalous infestation.  Many researchers focused on seeking source problem. In \cite{zarzhitsky2005swarms}, the authors used mobile robots  to measure diffusion source with gradient climbing  method. Parameter estimation algorithm \cite{demetriou2006power} was also used to identify a moving diffusion source. Recently, with the rapid development of unmanned aerial vehicle (UAV), multiple UAV are used for source seeking \cite{han2014multiple}. However, there is not enough information for controlling a diffusion process only by knowing its source.  Qiang Du et al. \cite{du1999centroidal} proposed Centroidal Voronoi Tessellations in coverage control, and then CVT method was extended to a diffusion and spray control \cite{chen2007optimal}. Here, inspired by the properties of less time with higher efficiency while using UAV, we try to use low-cost UAVs as cropdusters/actuators to control anomalous diffusion of pest spreading.

It can be easily seen that detecting and controlling a diffusion process will be viewed as an optimal sensor/actuator location problem in a distributed parameter system. Basically, CVT algorithm provides a non-model-based method for coverage and diffusion control by using a group of robots. Therefore, people use CVT to generate some desired actuator positions.

This paper tries to solve the problem how to locate a group of UAVs to sense and control anomalous diffusion process of pest infestation. What we most concern here is the minimal impact to the soil and agriculture caused by pesticide and pest. The pest has severe negative impact on agricultural industry, but the pesticide may  also have bad effect on the soil. So, the main objective for us is to deploy the UAVs and control the pest diffusion in an optimal way to minimize any negative impact on the crops and also the soil. That is to say, the pesticide should be released in such a way that the diffusion of the pest is bounded so that the heavily affected area is kept as small as possible. These all depend on how the UAVs positions are chosen, how they move and what control law is used to release the pesticide.

Motivated by real-time pest management by using unmanned cropdusters \cite{stark2013optimal} and the application of CVT in optimal placement of resource \cite{du1999centroidal} , we proposed a practical algorithm based on CVT to solve the problem of actuator motion planning to spray the pest. In our experiment, the pest/pollution density is
given by the sensors that cover the area and form a mesh. Based on \texttt{Diff-MAS2D} \cite{liang2004diff}, we develop a new simulation platform \texttt{FO-Diff-MAS2D} for  fractional diffusion  system measurement and control with mobile sensors and mobile actuators. Our proposed algorithm has been implemented on \texttt{FO-Diff-MAS2D}. Simulation results access the effectiveness of our algorithm as well as the role of fractional order in the overall control performance changes.

The remainder of this paper is organized as follows. In Section \ref{sec2}, the models of anomalous diffusion process and the corresponding control problems are formulated.
In Section \ref{sec3}, Centroidal Voronoi Tessellations (CVT) based  optimal actuator location algorithm is briefly introduced.  The main features of  our simulation platform \texttt{FO-Diff-MAS2D} for fractional diffusion  system measurement and control  are introduced in Section
\ref{sec4}.  In Section \ref{sec5}, we imply simulation experiments to show the effectiveness of our proposed methods. Finally, we conclude the paper and give
future research efforts in last section.
\section{Mathematical models and problem formulation}
\label{sec2}
In this section, we give two anomalous diffusion models to describe infestation problem. The two models are described by time fractional diffusion equation and space fractional diffusion, respectively.

Suppose that two anomalous diffusion processes evolve  in a convex polytope $\Omega:~\Omega\in R^2$. The time domain is $t\geq0$.
Let $\rho(x,y): \Omega \rightarrow R_+$ represent the pest density over $\Omega$. The corresponding dynamic process can be  modeled by using the following time fractional diffusion equation

\begin{equation}\label{eq1}
  \,_{C}D_{0,t}^\alpha \rho(x,y,t)=k_\alpha\left(\frac{\partial ^2\rho}{\partial
    x^2}+\frac{\partial ^2\rho}{\partial y^2}\right)+f_d(\rho,x,y,t)+f_c(\tilde{\rho},x,y,t),
\end{equation}
and space fractional diffusion equation below
\begin{equation}\label{eq2}
  \frac{\partial\rho(x,y,t) }{\partial t}=k_\beta\left(\frac{\partial ^\beta\rho}{\partial
    |x|^\beta}+\frac{\partial ^\beta\rho}{\partial |y|^\beta}\right)+f_d(\rho,x,y,t)+f_c(\tilde{\rho},x,y,t),
\end{equation}
where $\rho(x,y,t)$ is the pest density  to be controlled, $k_\alpha$ and $k_\beta$ are positive constants representing the diffusion rate, $f_d(\rho,x,y,t)$ is source, $\tilde{\rho}$ is the measured data of $\rho$ from the sensors, $f_c(\tilde{\rho},x,y,t)$ is
 the control input by mobile actuators to spray the pest, and its exact form depends on the closed-loop control law designed by the user based on certain control performance requirement.  $\,_{C}D_{0,t}^\alpha \rho$ is Caputo fractional derivative  of order $\alpha$~($0<\alpha\leq1$) defined by \cite{podlubny1998fractional}
 \begin{equation}\label{eq3}
 \,_{C}D_{0,t}^\alpha \rho=\left\{\begin{aligned}
&\frac{1}{\Gamma(1-\alpha)}\int_0^t (t-\tau)^{-\alpha}\rho'(x,y,\tau)d\tau,~~0<\alpha<1,\\
&\frac{\partial \rho}{\partial t},~~\alpha=1.
 \end{aligned}\right.
 \end{equation}

The operators $\frac{\partial^\beta \rho}{\partial |x|^\beta}$ and $\frac{\partial^\beta \rho}{\partial |y|^\beta}$ are Riesz fractional derivatives, which are, respectively defined as \cite{podlubny1998fractional}
\begin{equation}\label{eq4}
\frac{\partial^\beta \rho}{\partial |x|^\beta}=\left\{\begin{aligned}
&-c_\beta\bigg(\,_{RL}D_{a,x}^\beta \rho+\,_{RL}D_{x,b}^\beta \rho\bigg),~~1<\beta<2,\\
&\frac{\partial^2\rho}{\partial x^2},~~\beta=2.
\end{aligned}\right.
\end{equation}
where $\,_{RL}D_{a,x}^\beta$, $\,_{RL}D_{x,b}^\beta$ are left/right Riemann-Liouville derivatives for variable $x$ defined in \cite{podlubny1998fractional}

\begin{equation}\label{eq5}
\frac{\partial^\beta \rho}{\partial |y|^\beta}=\left\{\begin{aligned}
&-c_\beta\bigg(\,_{RL}D_{c,y}^\beta \rho+\,_{RL}D_{y,d}^\beta \rho\bigg),~~1<\beta<2,\\
&\frac{\partial^2\rho}{\partial y^2},~~\beta=2,
\end{aligned}\right.
\end{equation}
here $\,_{RL}D_{c,y}^\beta$, $\,_{RL}D_{y,d}^\beta$ are left/right Riemann-Liouville derivatives for variable $y$.  The parameter $c_\beta$ in \eqref{eq3} and \eqref{eq4} is $c_\beta=\frac{1}{2\cos(\beta\pi/2)}$.

  Denote by $P=(p_1,\cdots,p_n)$ be the location of $n$ actuators and  $|\cdot|$ be the Euclidean distance. Each actuator at position $p_i$ will receive information from sensors and then move and release the pesticide chemical by some control law. The objectives are:
\begin{itemize}
    \item Control the infestation diffusion  to a confined area.
    \item Cropdusting in a time optimal way while not making the area overdosed.
    \item Minimize the crops area that is heavily affected.
\end{itemize}
$n$ sensors will devide $\Omega$ into a set of $n$ polytopes $\mathcal{V}=\{V_1,\cdots,V_n\}$, $p_i\in V_i$, $V_i\cap V_j=\emptyset\  \mbox{for} \  i\neq j \ \mbox{and} \ \cup_{i=1}^{n}\bar{V_i}=\bar{\Omega}$ ($\bar V_i=V_i\cup \partial V_i$ and $\bar \Omega=\Omega\cup \partial \Omega$). In order  to control the diffusion process and minimize the heavily affected area, the actuators should be close to those areas with high pest concentrations so that the pest can be killed timely and does not diffuse further. But confining all actuators very close to the pollution source is not a good strategy. To
decide the positions of the actuators, we consider the minimizing of the following cost function

\begin{equation}\label{eqncost}
    \mathcal{K}(P,\mathcal{V})=\sum_{i=1}^n\int_{V_i}\rho(q)|q-p_i|^2dq \ \mbox{for} \ q\in \Omega.
\end{equation}

 One can see that to minimize $\mathcal{K}$, the distance $|q-p_i|$ should be small when the pest density $\rho(q)$ is big. It is the density function $\rho(q)$ that determines the optimal positions of the actuators. A necessary condition for $\mathcal{K}$ to be minimized is that $\{p_i,V_i\}_{i=1}^{k}$ is a CVT of $\Omega$ \cite{ju2002probabilistic}. Our algorithm is based on a discrete version of (\ref{eqncost}) and the density information comes from the measurements of the static, low-cost sensors.

\section{optimal actuator algorithm}
\label{sec3}
In this section, we give the algorithm to compute the locations of actuators by using CVT. Analogue to  \cite{chen2007optimal, chen2006optimal, chen2005actuation, chen2005optimal}, we also use Lloy's method to determine CVT. The algorithm is described as follows
\cite{ju2002probabilistic}:

Give a region $\Omega$, a density function $\rho(x)$ defined for all $x\in \bar{\Omega}$, and a positive
integer $k$
\begin{enumerate}
    \item select an initial set of $k$ points $\{z_i\}^k_{i=1}$ as the generators,
    \item construct the Voronoi sets $\{V_i\}^k_{i=1}$ associated with generators $\{z_i\}^k_{i=1}$,
    \item determine the mass centroids of the Voronoi sets $\{V_i\}^k_{i=1}$; these centroids form the new
    set of points $\{z_i\}^k_{i=1}$,
    \item If the new points meet some convergence criterion, terminate; other wise, return to step 2).
\end{enumerate}

 We assume that an actuator can communicate with the sensors and other actuators within radius $R_i$. $R_i$  can be changed. Here we introduce a distributed algorithm from  \cite{chen2007optimal} with mild modification. At the first execution of step 2) in the above Lloy's algorithm, each UAV will do the following:
\begin{enumerate}
    \item Assign its detection range $R_i$ with a small initial value, detect all its neighboring sensors with radius $R_i$.
    \item Construct its own Voronoi cell within the radius $R_i$.
    \item For every sensor $q_i$, compute $d_i=\mbox{max}|q_i-p_i|$.
    \item If $R_i > 2\times d_i $, stop. Otherwise set $R_i=2 \times R_i$, go to step 2).
\end{enumerate}
$R_i$ obtained at the first step will be used as the initial values for the following steps.  For example,  if successive 3 updates, the $R_i$ remains unchanged, then $R_i$ can be decreased, $R_i=R_i-\Delta r$ for some $\Delta r > 0$. This improvement on the algorithm from \cite{chen2007optimal, cortes2002coverage} helps to reduce the computation requirements.

The mobile actuators are treated as virtual particles and obey the second-order dynamical equation:
\begin{equation}
\label{control}
\ddot{p}_i=-k_p(p_i-\bar{p_i})-k_v\dot{p}_i.
\end{equation}
 The first term of \eqref{control} on the right hand side is  a proportional control law used as the force input to control the motion, where $\bar{p_i}$ is the computed mass centroid of the current Voronoi cell. The second term  is the viscous friction artificially introduced in \cite{howard2002mobile}. $k_v$ is the friction coefficient and $\dot{p}_i$ denotes the velocity of the actuator $i$.

\section{FO-DiffMAS-2D platform}
\label{sec4}
In this section, we introduce the simulation platform \texttt{FO-Diff-MAS2D}  for fractional diffusion  model system measurement and control with mobile sensors and mobile actuators.

The following two types of boundary conditions at boundary $\partial \Omega$ can be used in the problem formulation.
\begin{itemize}
\item Dirichlet boundary condition
  \begin{equation}
    \label{eq:diribd}
    \rho=C,
  \end{equation}
  where $C$ is a real constant.
\item Neumann boundary condition
  \begin{equation}
    \label{eq:neumbd}
    \frac{\partial \rho}{\partial n}=C_1+C_2\rho,
  \end{equation}
  where $C_1$ and $C_2$ are two real constants; $n$ stands for the outward direction.
\end{itemize}

\texttt{FO-Diff-MAS2D} uses finite difference method  to discretize the spatial  derivative  in \eqref{eq1}, and fractional central difference \cite{ortigueira2008fractional} to approximate space fractional derivative in \eqref{eq2}, then leaves the time domain integration to Matlab/Simulink. Specifically, \texttt{FO-Diff-MAS2D} is used to solve a two dimensional fractional diffusion equation. See Appendix for the numerical verifications. As an extension of \texttt{Diff-MAS2D} \cite{liang2004diff}, the main features of \texttt{FO-Diff-MAS2D} are listed as
follows:
\begin{itemize}
\item Sensors and actuators can be collocated or
  non-collocated.
\item Disturbances can be movable and time-varying. \item The mobility platform dynamics of sensors and
actuators can be modeled as either first order  or second order.
\item Movement of sensors and actuators can be open-loop  or closed-loop.
\item Arbitrary control algorithms can be applied in $f_c(\tilde \rho(x,y,t),x,y,t)$.
\end{itemize}
\section{Simulation experiments}
\label{sec5}
In this section, we consider two different fractional diffusion models to describe the anomalous diffusion infestation of crops.
\texttt{FO-Diff-MAS2D} is used as the simulation platform for our realization. The area concerned is given by $\Omega=\{(x,y)|0\leq x\leq 1,~0\leq y\leq 1\}$.

\textbf{Example I}
The complex transport dynamic system with control input is modelled as the following time fractional diffusion equation

\begin{equation}
  \label{eq:parapde}
  \begin{aligned}
  \,_{C}D_{0,t}^\alpha \rho(x,y,t)&=0.01\left(\frac{\partial ^2\rho(x,y,t)}{\partial
    x^2}+\frac{\partial ^2\rho(x,y,t)}{\partial y^2}\right)\\
    &\qquad+f_c(x,y,t)+f_d(x,y,t),
    \end{aligned}
\end{equation}

with homogeneous Neumann boundary, given by
\[\frac{\partial u}{\partial n}=0.\]

The pest source is modeled as a point disturbance $f_d$ to the fractional diffusion system
(\ref{eq:parapde}) with its position at $(0.8,0.2)$ and \[f_d(t)=20e^{-t}|_{(x=0.8,y=0.2)}.\]

\begin{figure}[]
\centerline{\epsfig{file=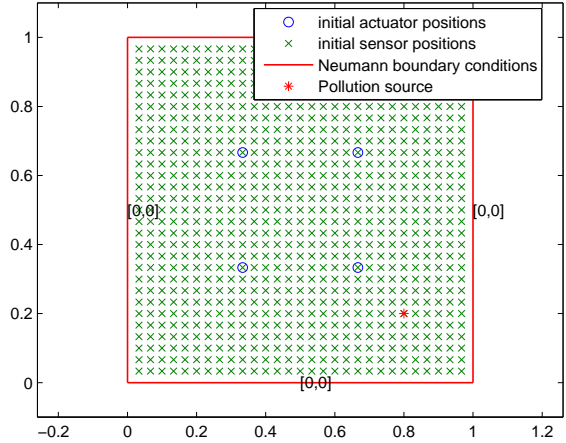,width=9cm}} \caption{Initial layout of actuators, sensors and obstacle.}
\label{initial}
\end{figure}

The pest source begins to move  at $t=0$ to the area $\Omega$, 4 actuators which can release the pesticide are deployed with initial
positions at $(0.33, 0.33)$, $(0.33, 0.66)$, $(0.66, 0.33),(0.66, 0.66)$, respectively. There are $29\times29$ sensors evenly distributed in a square area $(0,1)^2$, and they form a mesh over the area. In our simulation, we suppose that once deployed, the sensors remain static. Figure~\ref{initial} shows the initial positions of the actuators, the positions of the sensors and the position of the pollution source.

\begin{figure}[!htb]
\centering \mbox{\subfigure[Integer oder diffusion model] {\epsfxsize 60mm\epsffile{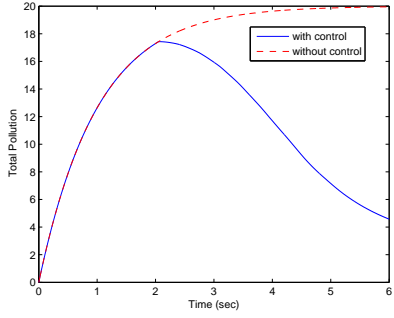}\label{integer}} }
\mbox{\subfigure[Fractional order diffusion model~($\alpha=0.99$)] {\epsfxsize 60mm\epsffile{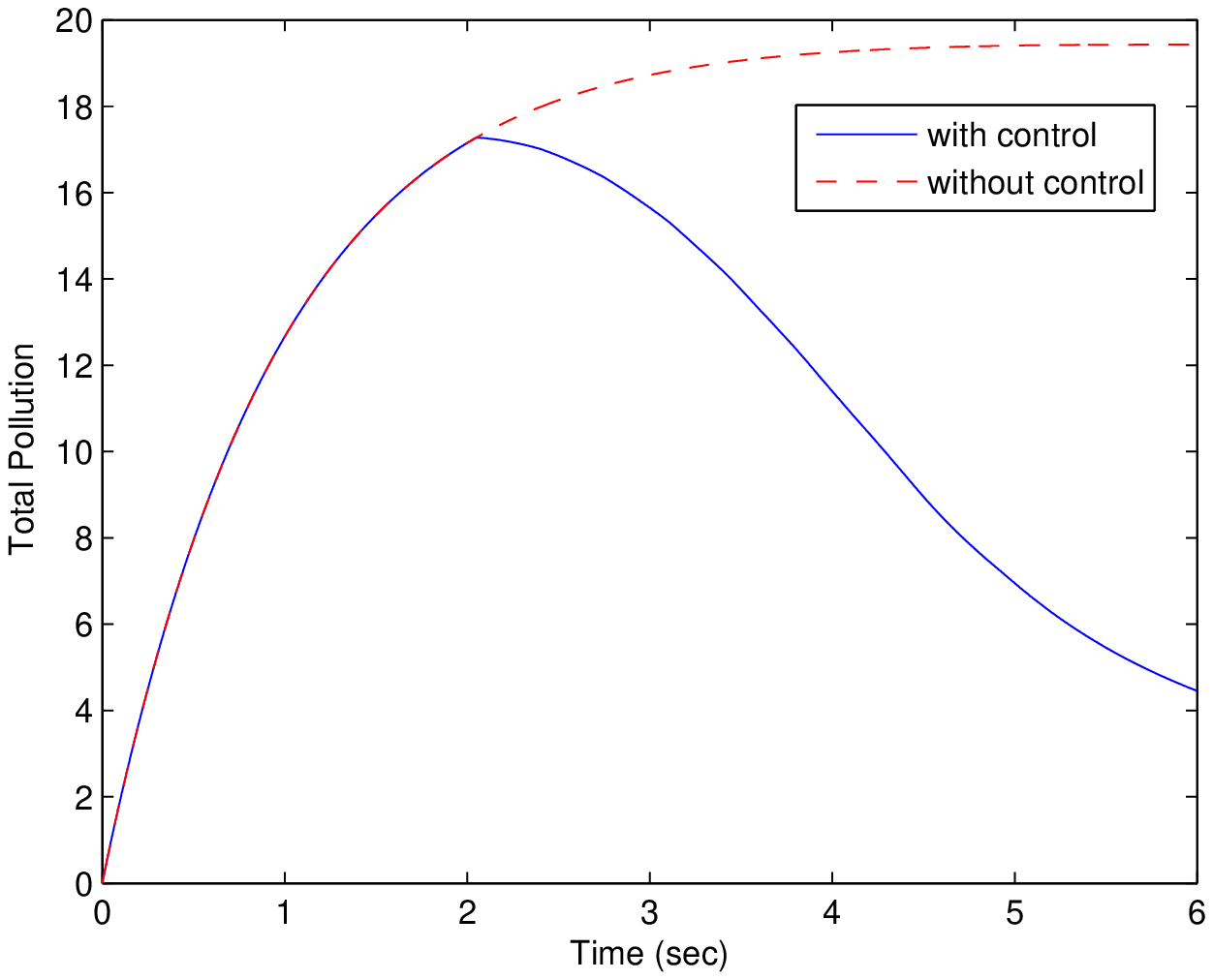}\label{.99}}}
 \caption{Evolution of the amount of pollutants } \label{comp}
\end{figure}

To check  the validity of time fractional diffusion model, we give the simulation result in \cite{chen2007optimal} for comparision, the results are displayed in Figure~\ref{comp}. The y axis is the sum of the sensors measurements. It is clear that the result  by using time fractional diffusion model is  in accordance with integer order diffusion model when  order $\alpha \approx 1$.

\begin{figure}[!htpb]
\centerline{\epsfig{file=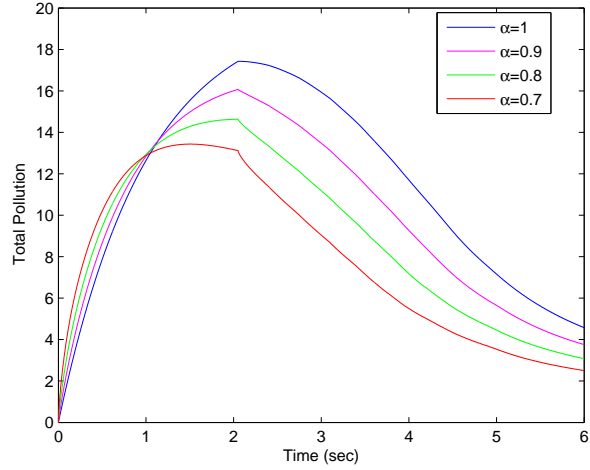,width=9cm}} \caption{Comparision different evolution  with different values of $\alpha$}
\label{compare}
\end{figure}

 In order to illustrate the different diffusion behavior and control effectiveness with different time fractional derivative order, we choose $\alpha=0.7$,$0.8$, $0.9$, $1$, respectively. The simulation resluts are shown in Figure.~\ref{compare},  but the system  is divergent when $\alpha<0.7$,  $\alpha=0.7$ is the best derivative order in equation \eqref{eq:parapde}. One can see that the peak of the pollution  becomes lower as $\alpha$ decreases. It can seen that  the pest diffusion exhibit some behavior of subdiffusion.

\begin{figure}[!htpb]
\centerline{\epsfig{file=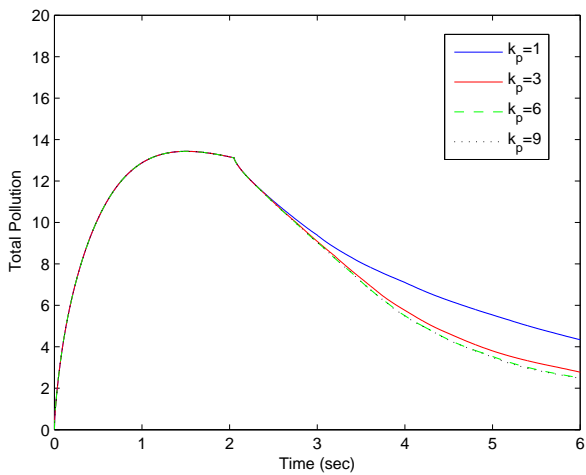,width=9cm}} \caption{Evolution of the amount of pollutants with different control law~($\alpha=0.7$)}
\label{diffk}
\end{figure}

  We now fix $\alpha=0.7$ as the optimal time derivative order in  equation \eqref{eq:parapde}. First, for designing the actuator motion control law, the viscous coefficient is fixed by $k_v=1$, in order to obtain optimal control strategy, letting $k_p$ change from 1 to 9, as shown in Figure~\ref{diffk}, $k_p=6$ is the best gain. Then, we give the following control input
  \begin{equation}\label{conlaw}
    \ddot{p}_i=-6(p_i-\bar{p_i})-\dot{p}_i.
    \end{equation}
We choose the simulation time to $t=6$s. And the step size is chosen as $\Delta t=0.002$s. The actuator recomputes its desired position every $0.1$s. To show how the actuators can control the diffusion of the pests, the UAVs begin to react at $t=0.4$s. The system evolves under the effects of diffusion of
pests and diffusion of pesticide released by UAVs.

\begin{figure}[!htpb]
\centerline{\epsfig{file=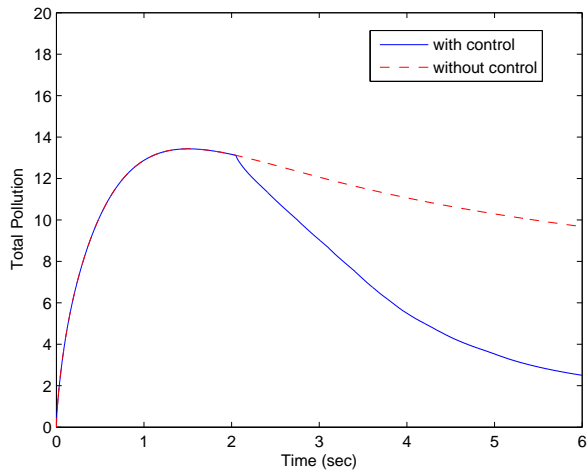,width=9cm}} \caption{Evolution of the amount of pollutants with $\alpha=0.7$}
\label{evo}
\end{figure}

In Figure~\ref{evo}, the y axis is the sum of the sensors measurements. It shows that the amount of pests decreases to $19\%$ of its peak value at the end of the simulation. And the decreasing process is monotonic. The evolution of the amount of pests without control is also shown in Figure~\ref{evo}. we can tell that the pest infestation of crops has been well controlled.

\begin{figure}[!htb]
\centering \mbox{\subfigure[$t=1.0$s.] {\epsfxsize 60mm\epsffile{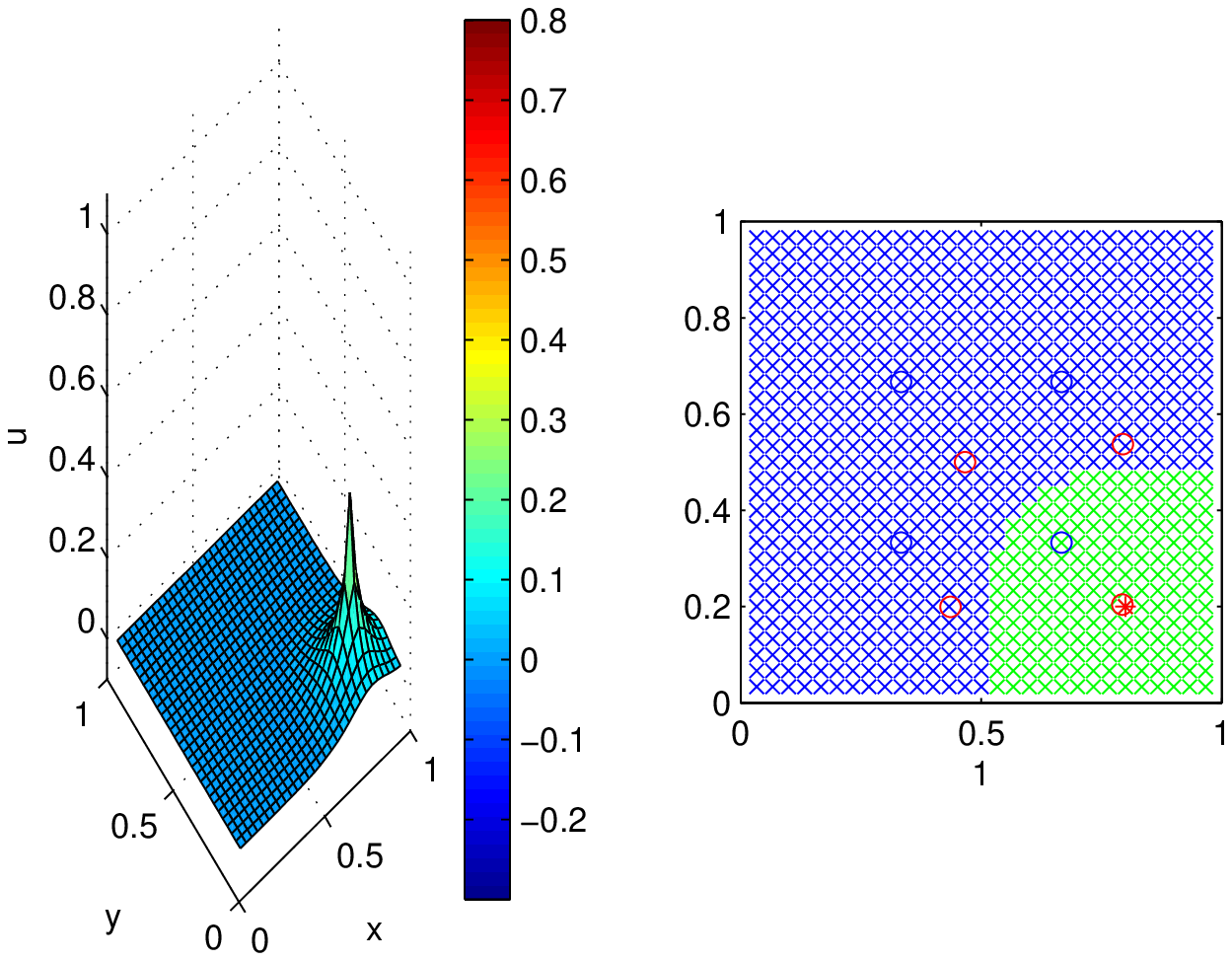}\label{10}} }
\mbox{\subfigure[ $t=2.0$s] {\epsfxsize 60mm\epsffile{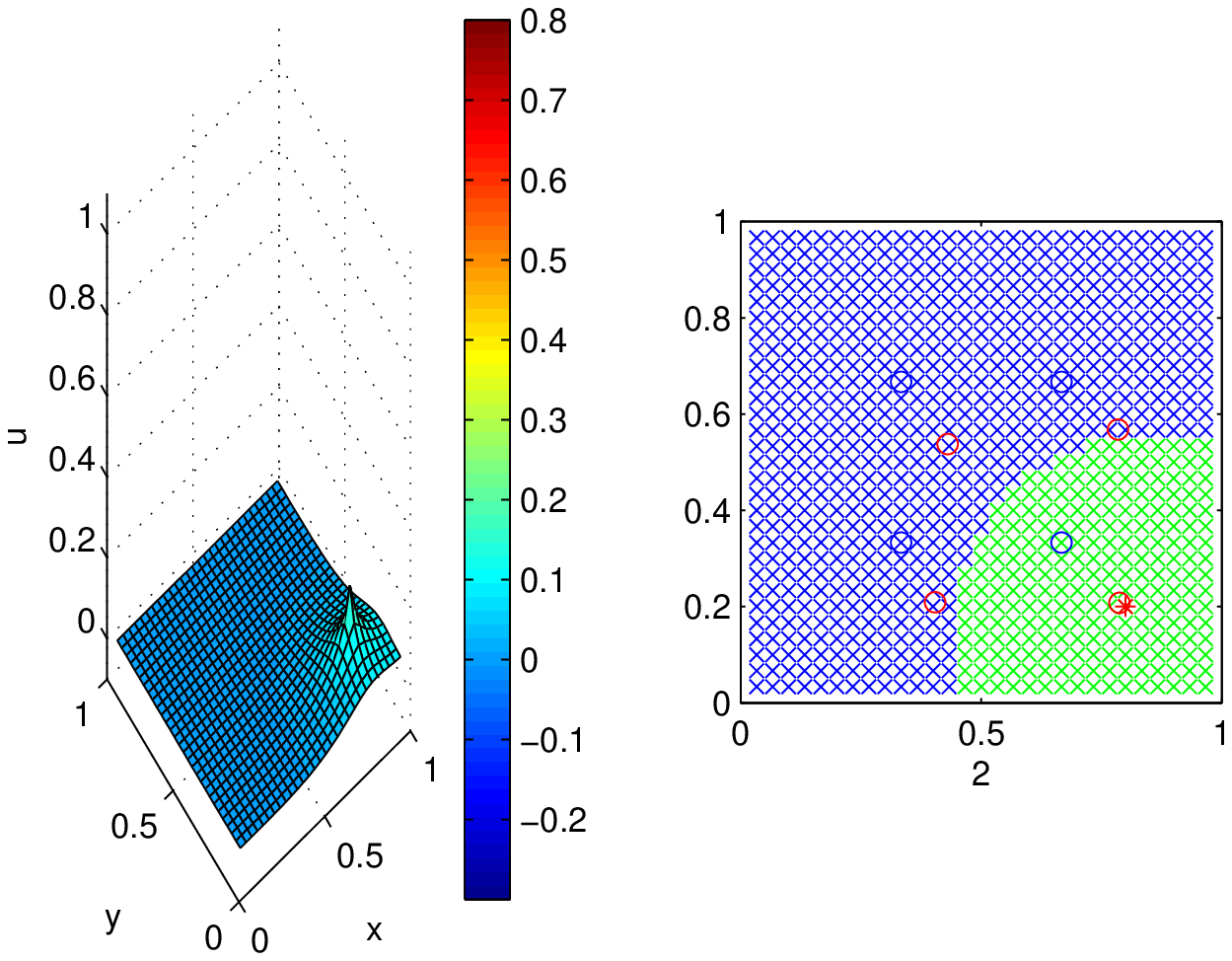}\label{20}}}
\mbox{\subfigure[$t=3.0$s] {\epsfxsize 60mm\epsffile{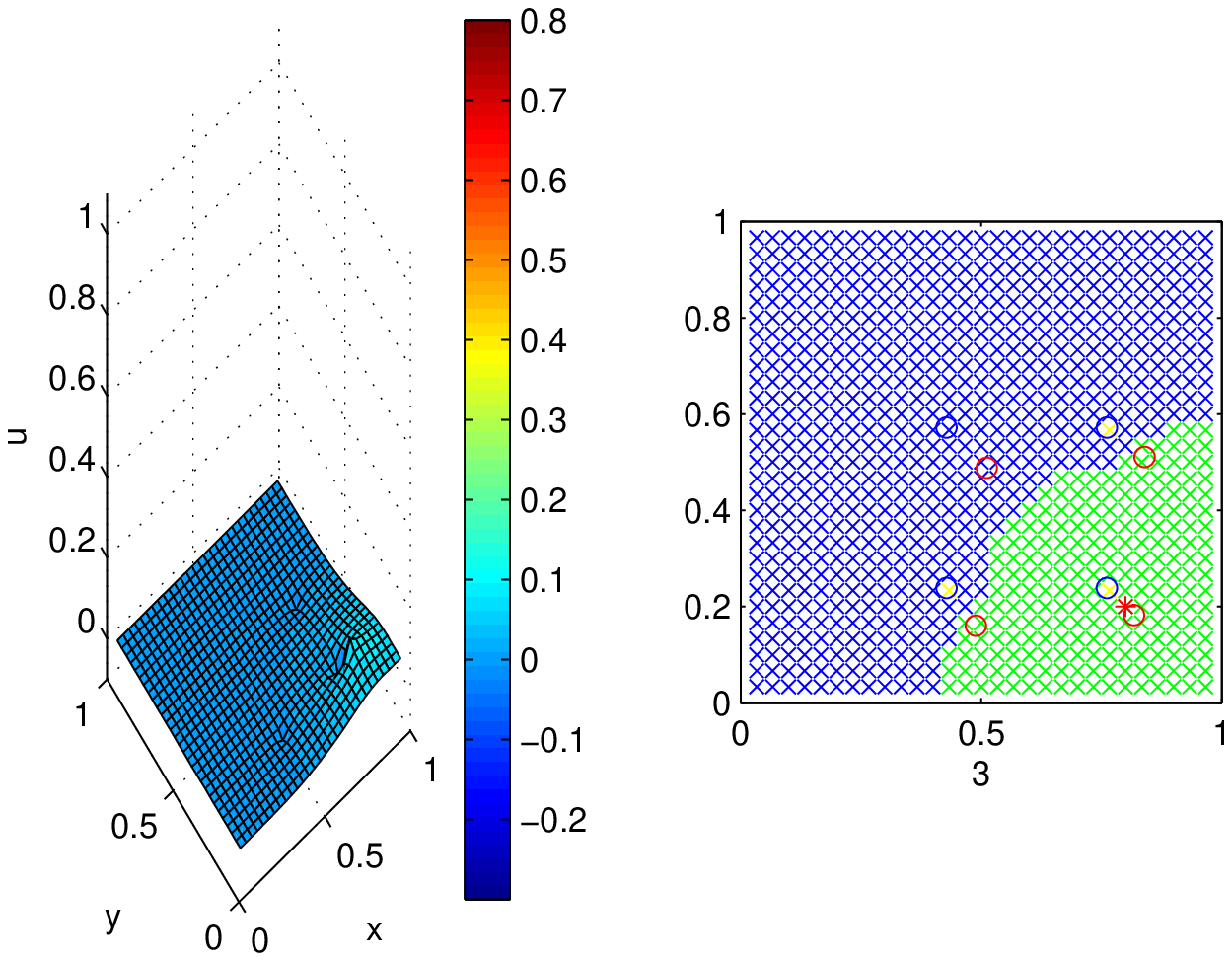}\label{30}}}
\mbox{\subfigure[ $t=4.0$s] {\epsfxsize 60mm\epsffile{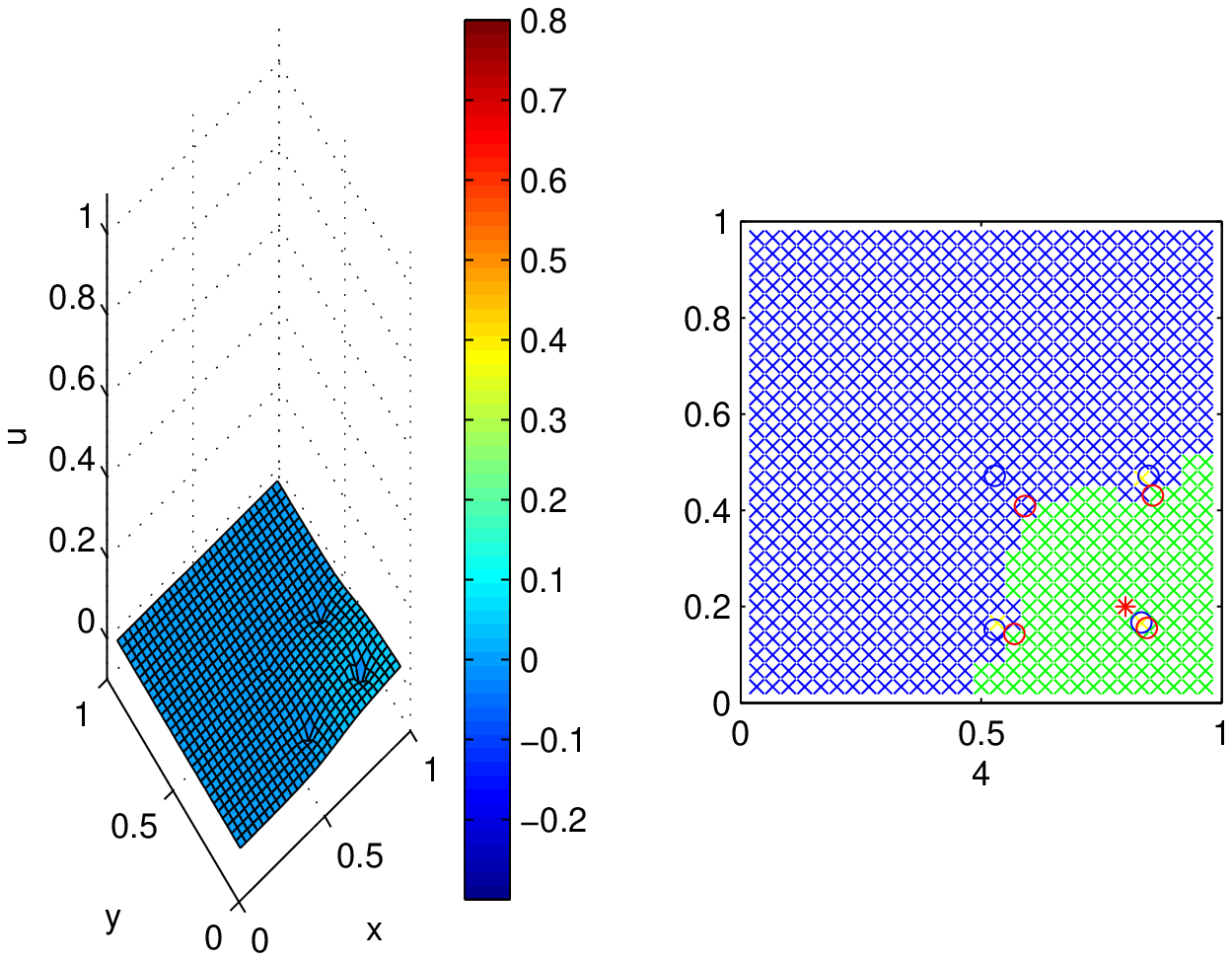}\label{40}}}
\mbox{\subfigure[$t=5.0$s] {\epsfxsize 60mm\epsffile{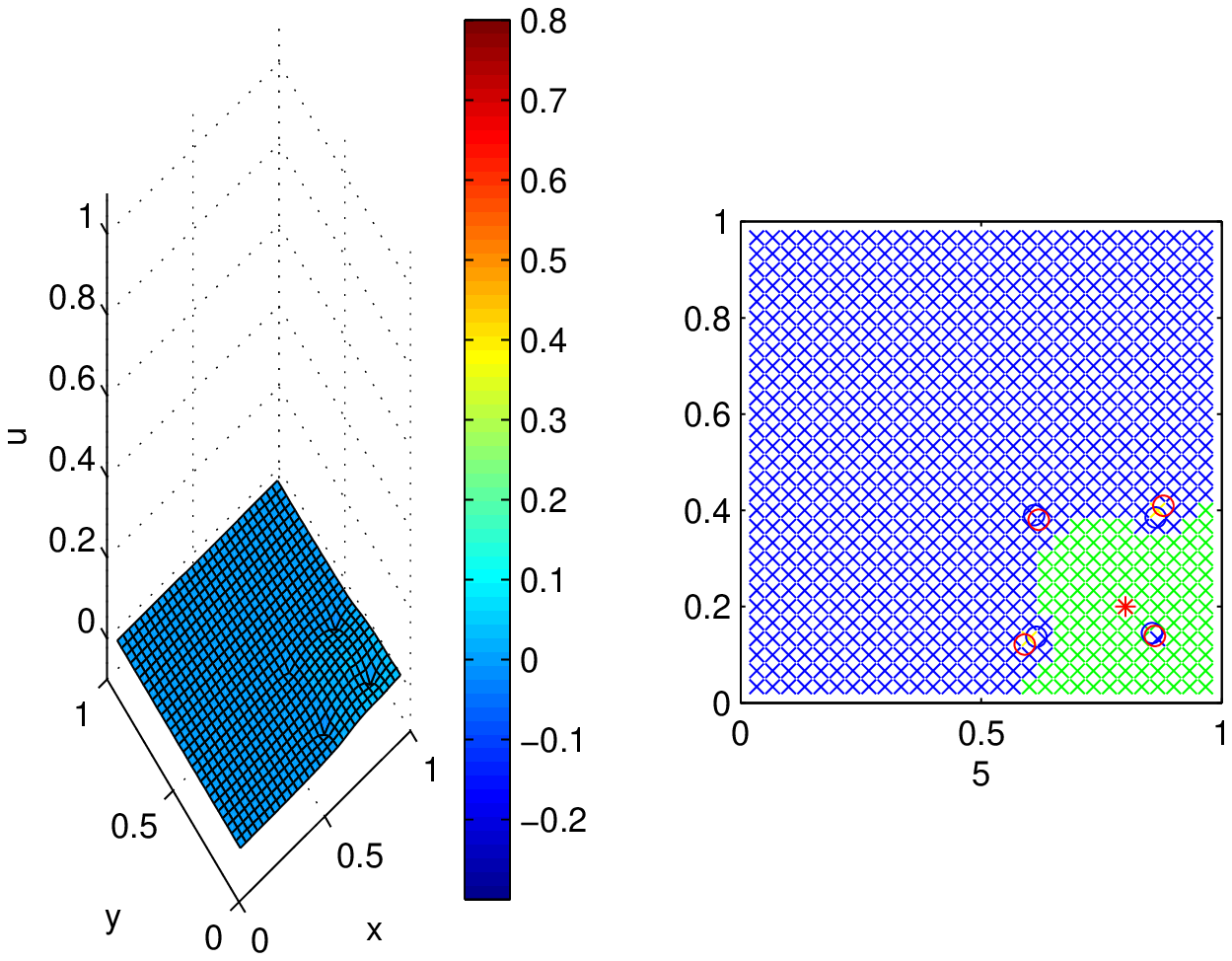}\label{50}}}
\mbox{\subfigure[$t=5.5$s] {\epsfxsize 60mm\epsffile{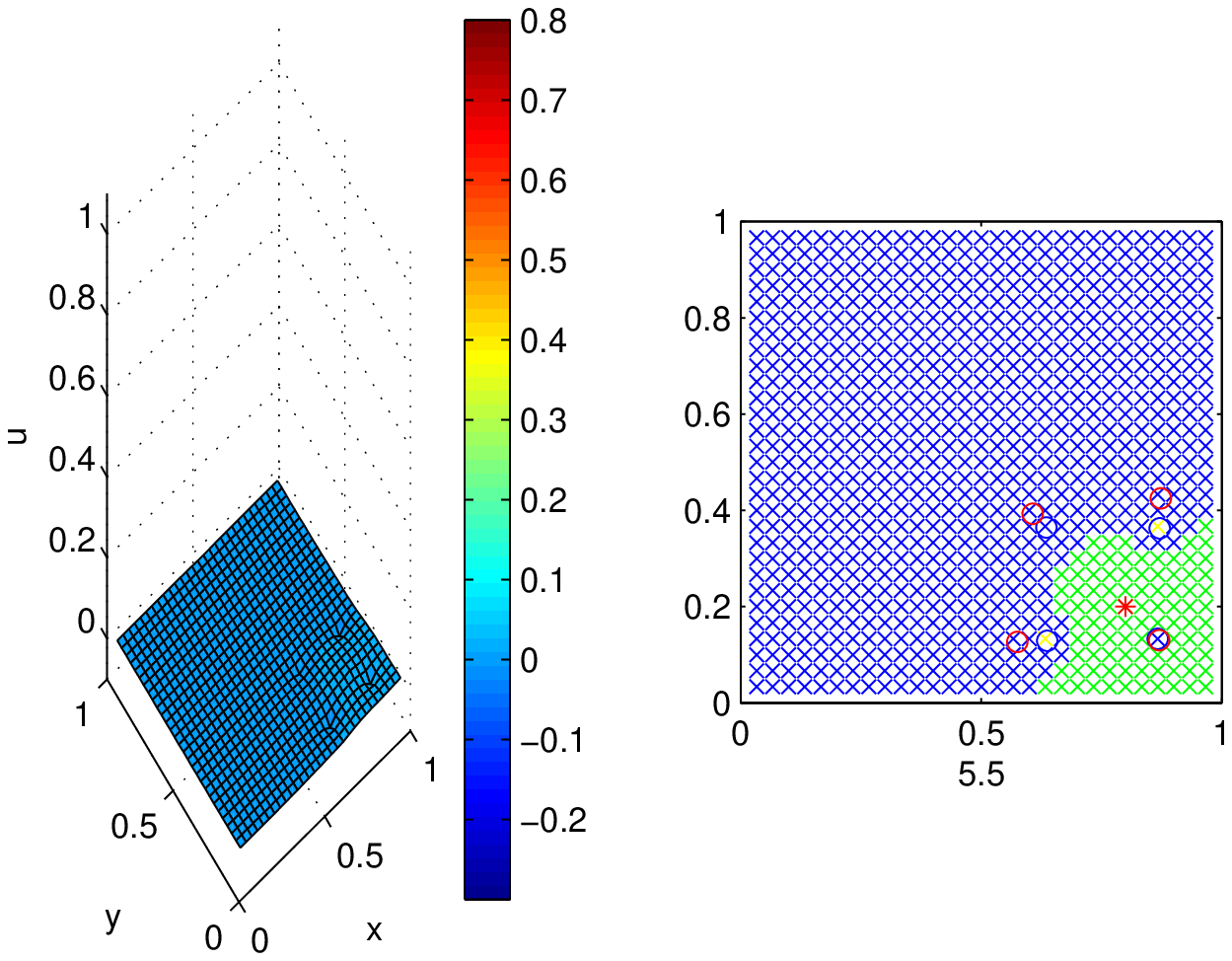}\label{55}}}
 \caption{Evolution of $\rho(x,y,t)$, mobile robots positions and desired positions} \label{v}
\end{figure}

Figure~\ref{v} shows the evolution of the diffusion process with 4 actuators controlling one pest source diffusing through the region. The blue circles represent the actuator positions. The red circles resprent the desired actuator positions using CVT.

It can be seen that the pollution density is becoming lower when the actuators are approaching their desired positions. It can be concluded that the pollution has been well controlled by equation \eqref{eq:parapde} with  fractional diffusion equation combining actuator control law \eqref{conlaw}.

\begin{figure}[]
\centerline{\epsfig{file=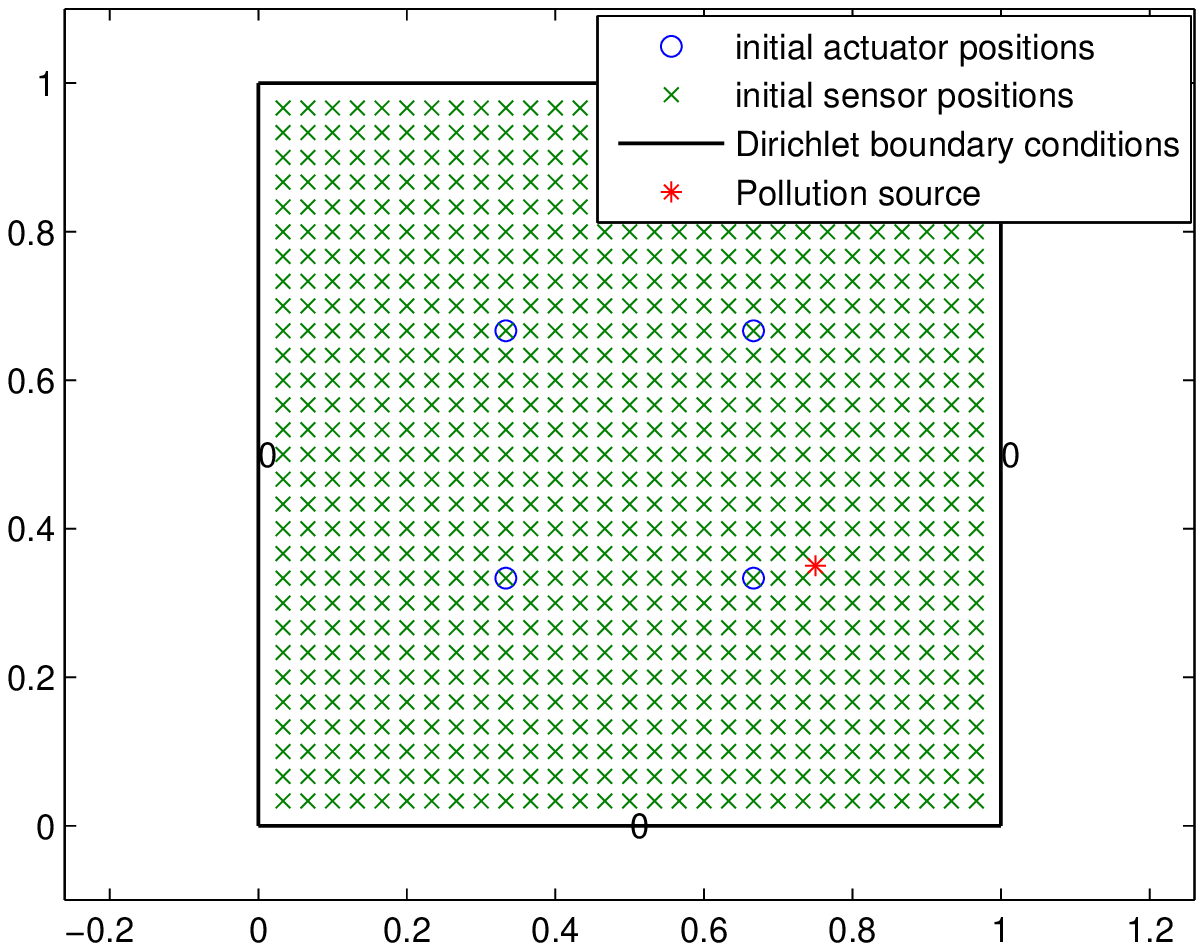,width=9cm}} \caption{Initial layout of actuators, sensors and obstacle.}
\label{intial}
\end{figure}

\textbf{Example II} The system is modelled by the following space fractional diffusion equation
\begin{equation}\label{ries}
\frac{\partial \rho(x,y,t)}{\partial t}=0.01\left(\frac{\partial ^\beta\rho}{\partial
    |x|^\beta}+\frac{\partial ^\beta\rho}{\partial |y|^\beta}\right)+f_d(\rho,x,y,t)+f_c(\tilde{\rho},x,y,t),
\end{equation}
combining with the homogeneous Dirichlet boundary condition
\[\rho=0.\]
The pest source is modeled as a point disturbance $f_d$ to the fractional diffusion system
(\ref{ries}) with its position at $(0.75,0.35)$ and \[f_d(t)=20e^{-t}|_{(x=0.75,y=0.35)}.\]
The pest source begins to move  at $t=0$ to the area $\Omega$, 4 robots which can release the pesticide are deployed with initial positions at $(0.33, 0.33)$, $(0.33, 0.66),(0.66, 0.33),(0.66, 0.66)$, respectively. There are $29\times29$ sensors evenly distributed in a square area $(0,1)^2$, and they form a mesh over the area. In our simulation, we assume that once deployed, the sensors remain static. Figure~\ref{intial} shows the initial positions of the robots, the positions of the sensors and the position of the source.

\begin{figure}[!htpb]
\centerline{\epsfig{file=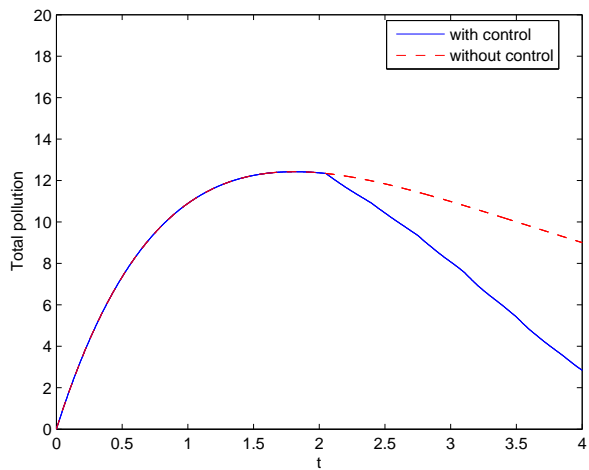,width=9cm}} \caption{Evolution of the amount of pollutants with $\beta=1.7$}
\label{evo2}
\end{figure}

In order to illustrate the different diffusion behavior and control effectiveness with different space fractional derivative order, we choose $\beta=2,~1.9, ~1.7,~1.6,~1.5$, respectively, and simulation time is $t=4$ s. The simulation resluts are shown in Figure~\ref{comp2}. From the figure, we can see that  $\beta=1.7$ is the best spatial derivative order in equation \eqref{ries} to model the space fractional diffusion process.
\begin{figure}[!htpb]
\centerline{\epsfig{file=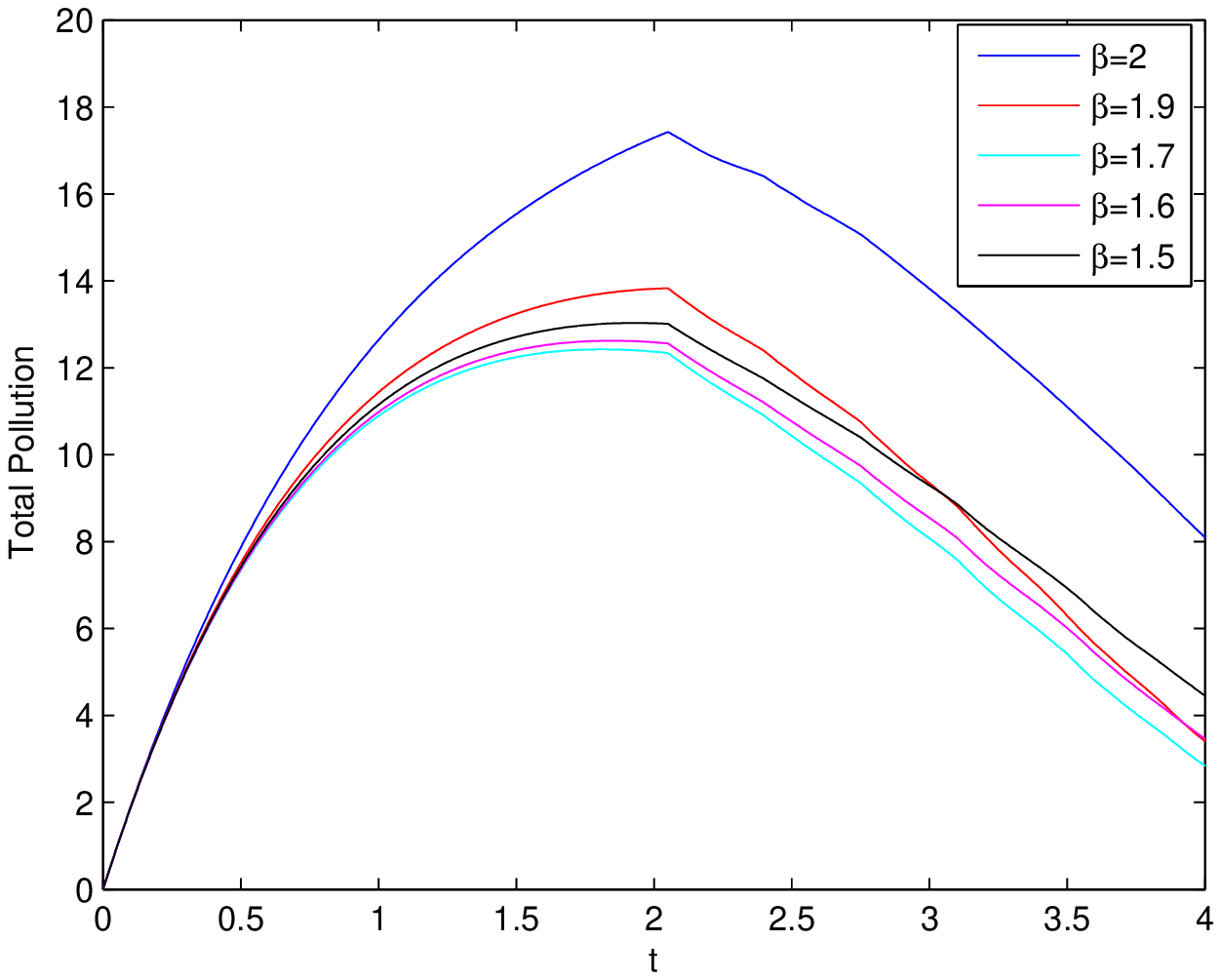,width=9cm}} \caption{Comparision different evolution  with different values of $\beta$}
\label{comp2}
\end{figure}
Thus, we fix $\beta=1.7$ in \eqref{ries}, and use the same control law in case I, the numerical simulation result is shown in Figure~\ref{evo2}. One can see that the space fractional diffusion of pest infestation is well controlled.

\section{Conclusions}
In this paper, we consider the optimal dynamic location problem for a group of UAVs which can release pesticide chemicals, known as  ``mobile actuator'' or ``cropdusters" for neutralizing a pest  diffusion process which are governed  by time  fractional order diffusion equation and space fractional diffusion equation. A new simulation platform has been built to prove the effectiveness of our algorithm and the problem of optimal dynamic motion of mobile actuators has been solved by CVT with better effectiveness. We also obtain the optimal value of derivative order for both time fractional order and space fractional order, when they are used to describe the corresponding anomalous diffusion process.  In the future, we will extend our UAV-based optimal crop-dusting problem by using fractional order control law.

\section*{APPENDIX}
To verify that FO-DiffMAS-2D works better  for sovling the anomalous diffusion problem, we use it to numerically solve the following two dimensional fractional diffusion equation. 

The matlab code used in the simulations can be downloaded from here: http://www.mathworks.com/matlabcentral
/fileexchange/48675-a-numerical-simulation-platform-for-the-control-of-anomalous-diffusion-process.

\textbf{Example 1}
Consider the following two dimensional time fractional diffusion equation
\begin{equation}
\begin{aligned}
\,_{C}D_{0,t}^\alpha u(x,y,t)&=0.01\bigg(\frac{\partial^2 u}{\partial x^2}+\frac{\partial^2u}{\partial y^2}\bigg)+f(x,y,t),\\
&\quad 0\leq x\leq 1,~0\leq y\leq 1,~t\geq 0,
\end{aligned}
\end{equation}
with initial and boundary conditions
\begin{equation}
u(x,y,0)=0,\quad 0<x<1,~0<y<1,
\end{equation}
\begin{equation}
\begin{aligned}
&u(0,y,t)=u(1,y,t)=0,\\
&u(x,0,t)=u(x,1,t)=0,
\end{aligned}
\end{equation}
where $$f(x,y,t)=\frac{2t^{2-\alpha}}{\Gamma(3-\alpha)}(x-x^2)(y-y^2)+0.02t^2(x-x^2+y-y^2).$$
The exact solution of the above equations is
$$u(x,y,t)=t^2 x(1-x)y(1-y).$$

Then, let $h_x=h_y=\frac{1}{20}$ as spatial step size, $\tau=\frac{1}{250}$ as time step length, the comparisons between numerical and exact solutions at time $t=0.5$ with different $\alpha$ are displayed in the followings
\begin{figure}[!htb]
\centering \mbox{\subfigure{\epsfxsize 60mm\epsffile{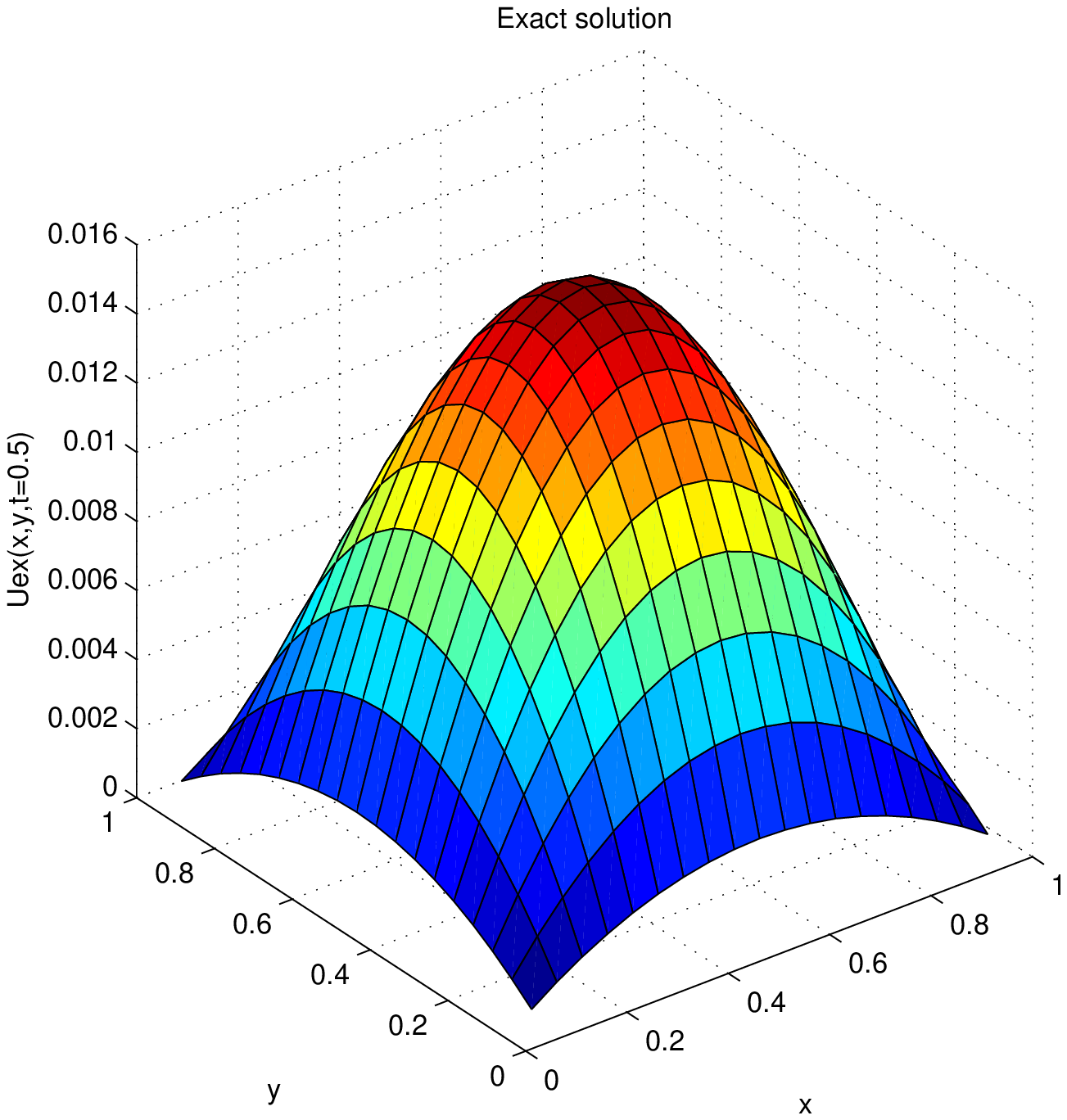}\label{exact}} }
\mbox{\subfigure {\epsfxsize 60mm\epsffile{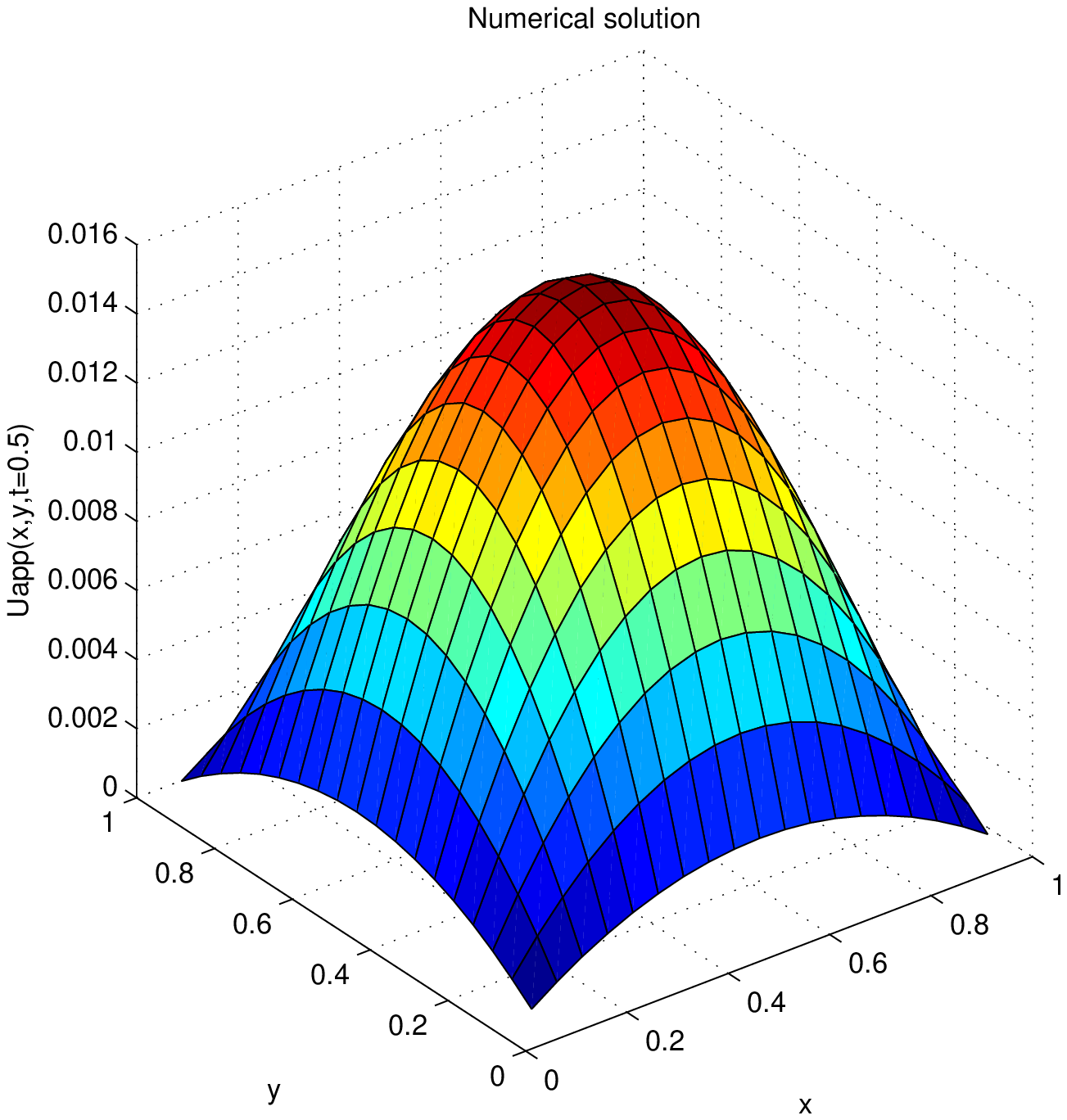}\label{1}}}
 \caption{Comparison of numerical and exact solution when $\alpha=0.6$} \label{comp1}
\end{figure}

\begin{figure}[!htb]
\centering \mbox{\subfigure{\epsfxsize 60mm\epsffile{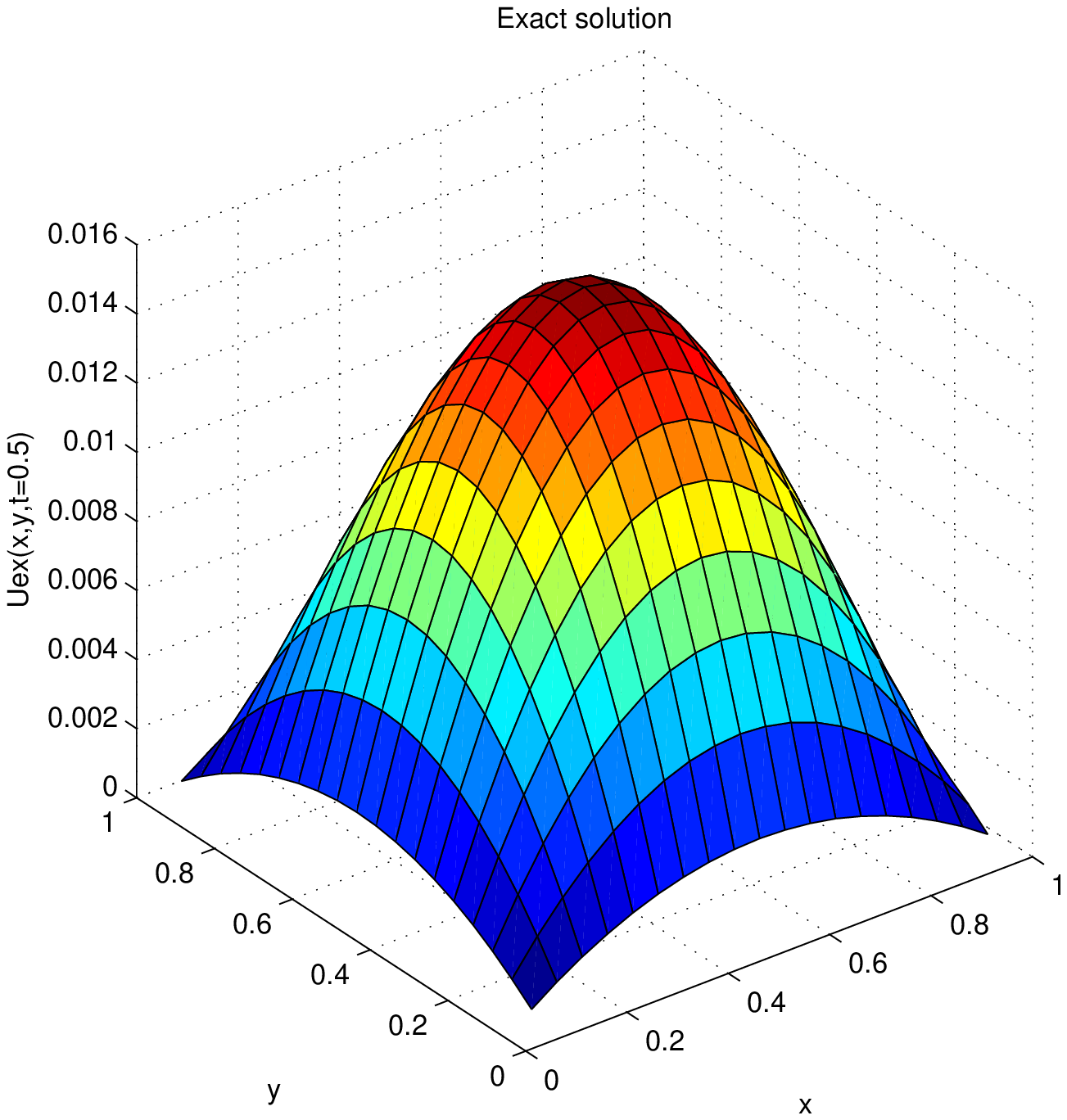}\label{exact}} }
\mbox{\subfigure {\epsfxsize 60mm\epsffile{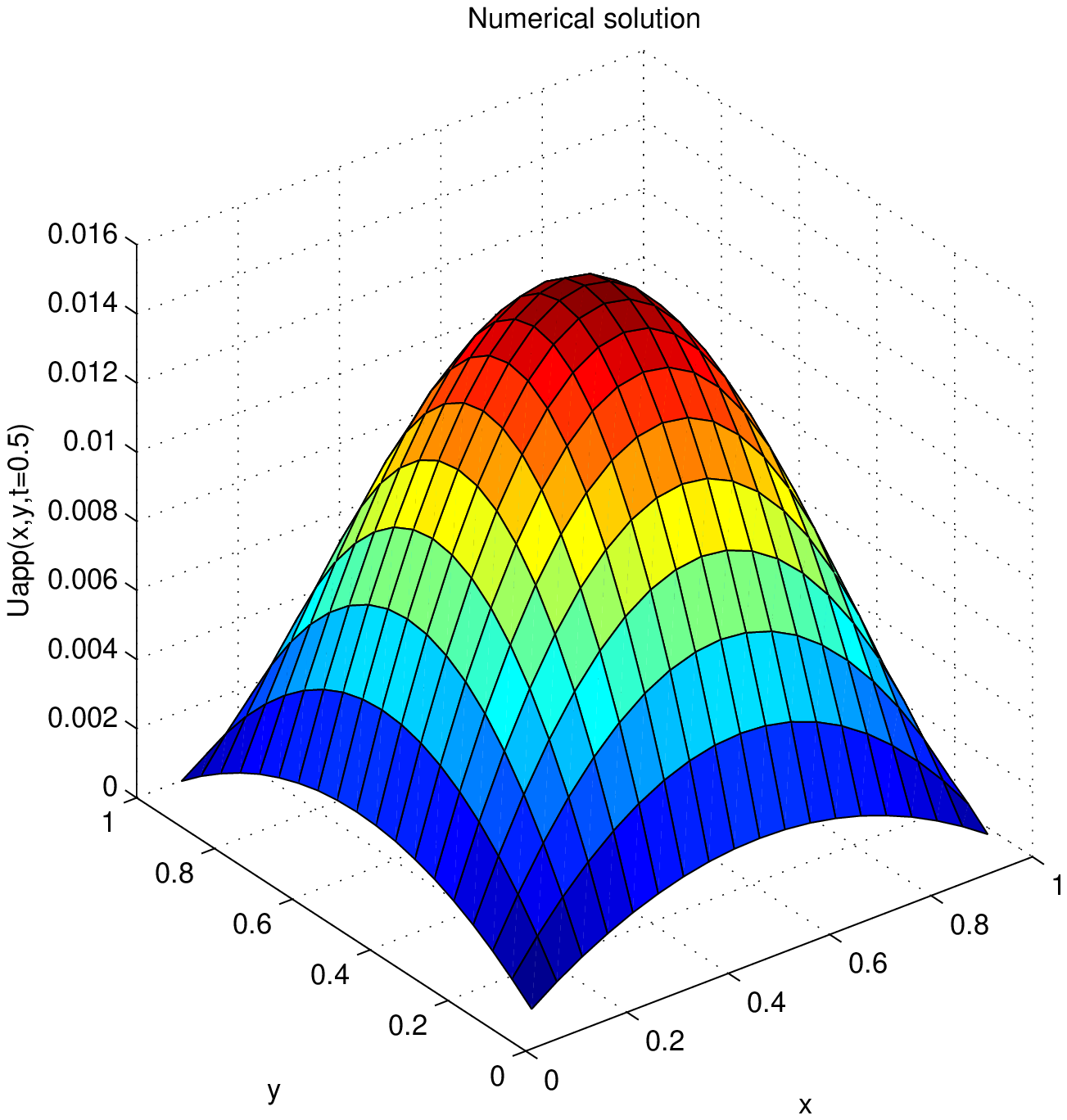}\label{1}}}
 \caption{Comparison of numerical and exact solution when $\alpha=0.7$} \label{comp2}
\end{figure}

\begin{figure}[!htb]
\centering \mbox{\subfigure{\epsfxsize 60mm\epsffile{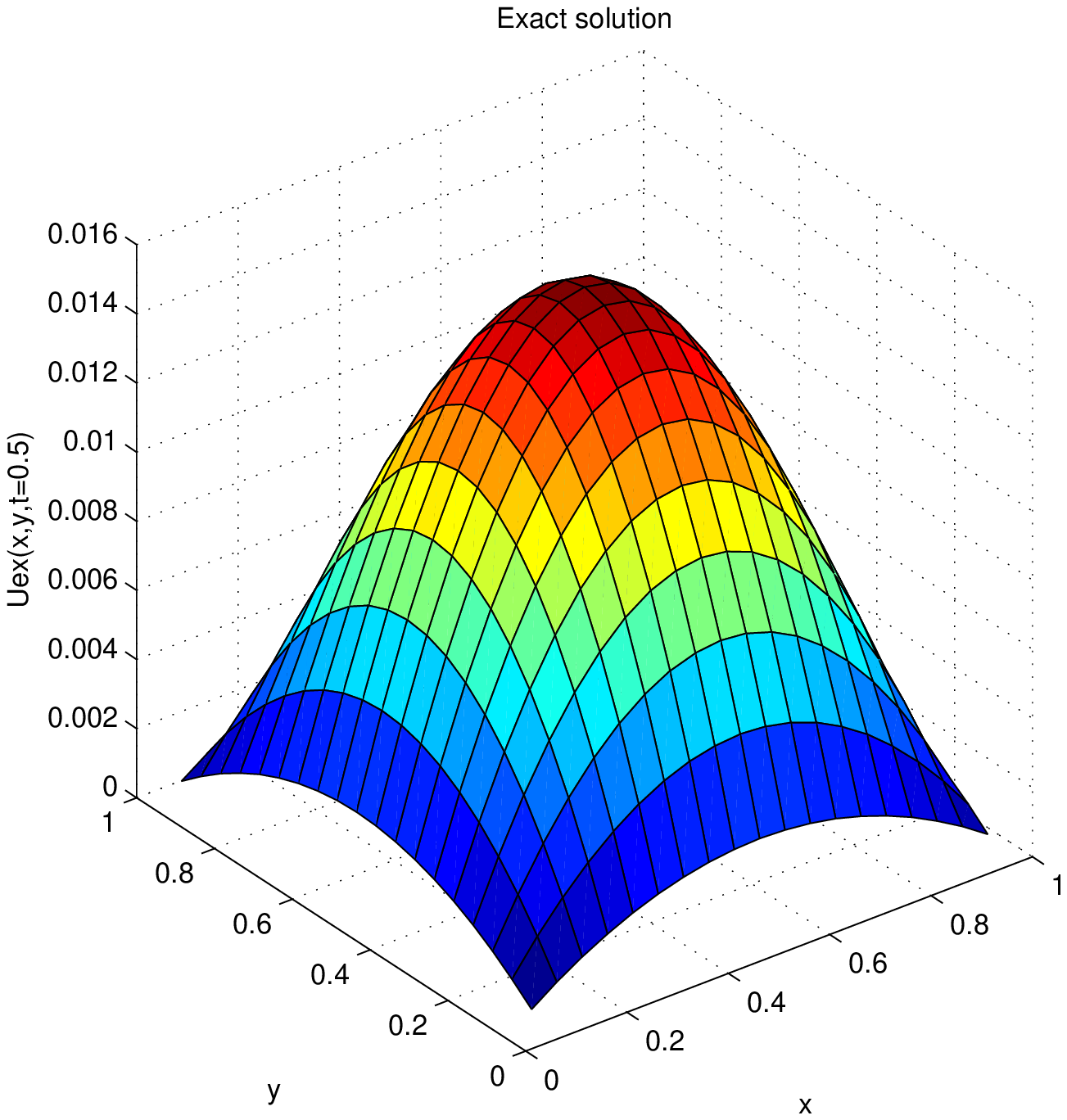}\label{exact}} }
\mbox{\subfigure {\epsfxsize 60mm\epsffile{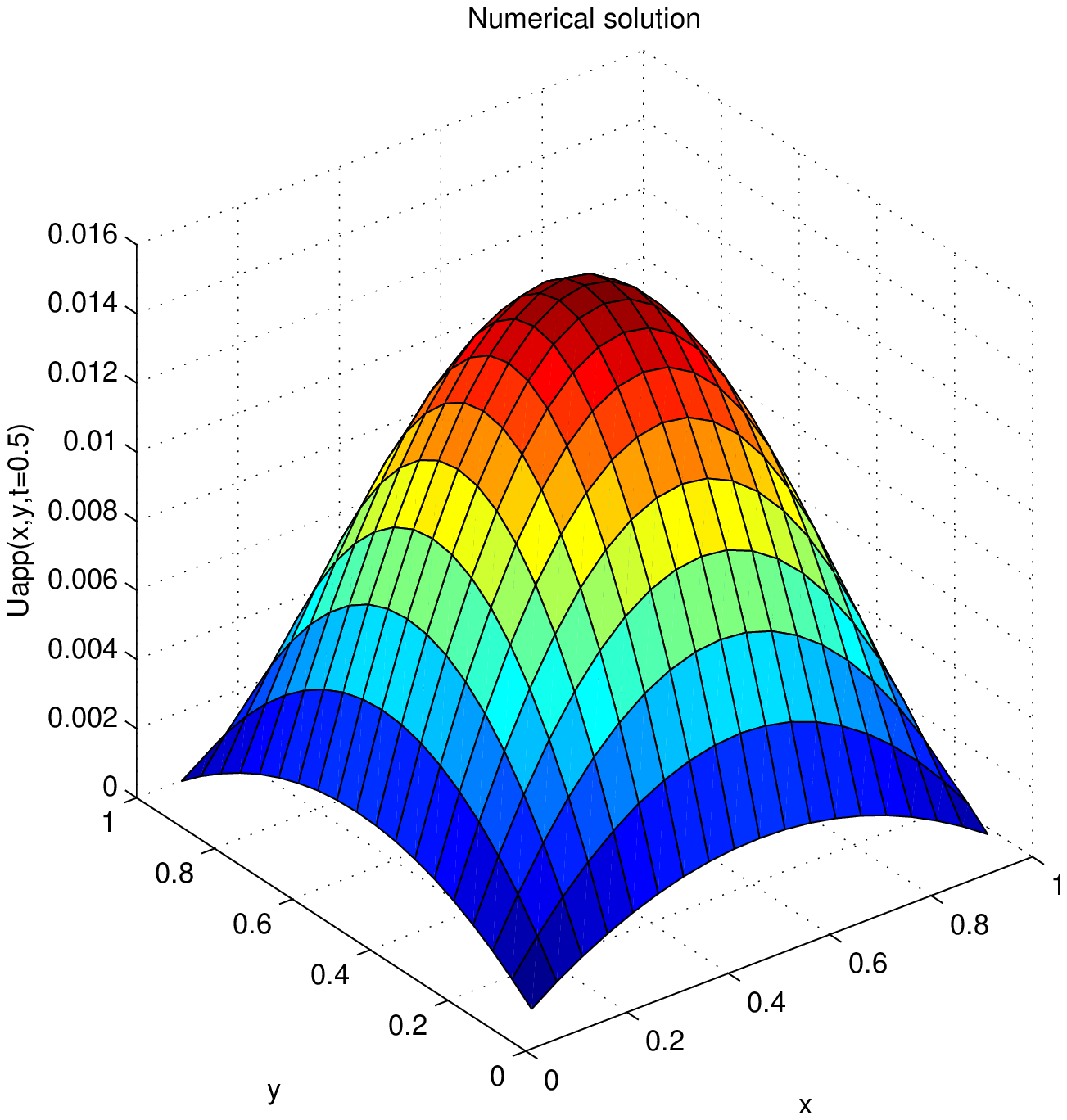}\label{1}}}
 \caption{Comparison of numerical and exact solution when $\alpha=0.8$} \label{comp3}
\end{figure}

\begin{figure}[!htb]
\centering \mbox{\subfigure{\epsfxsize 60mm\epsffile{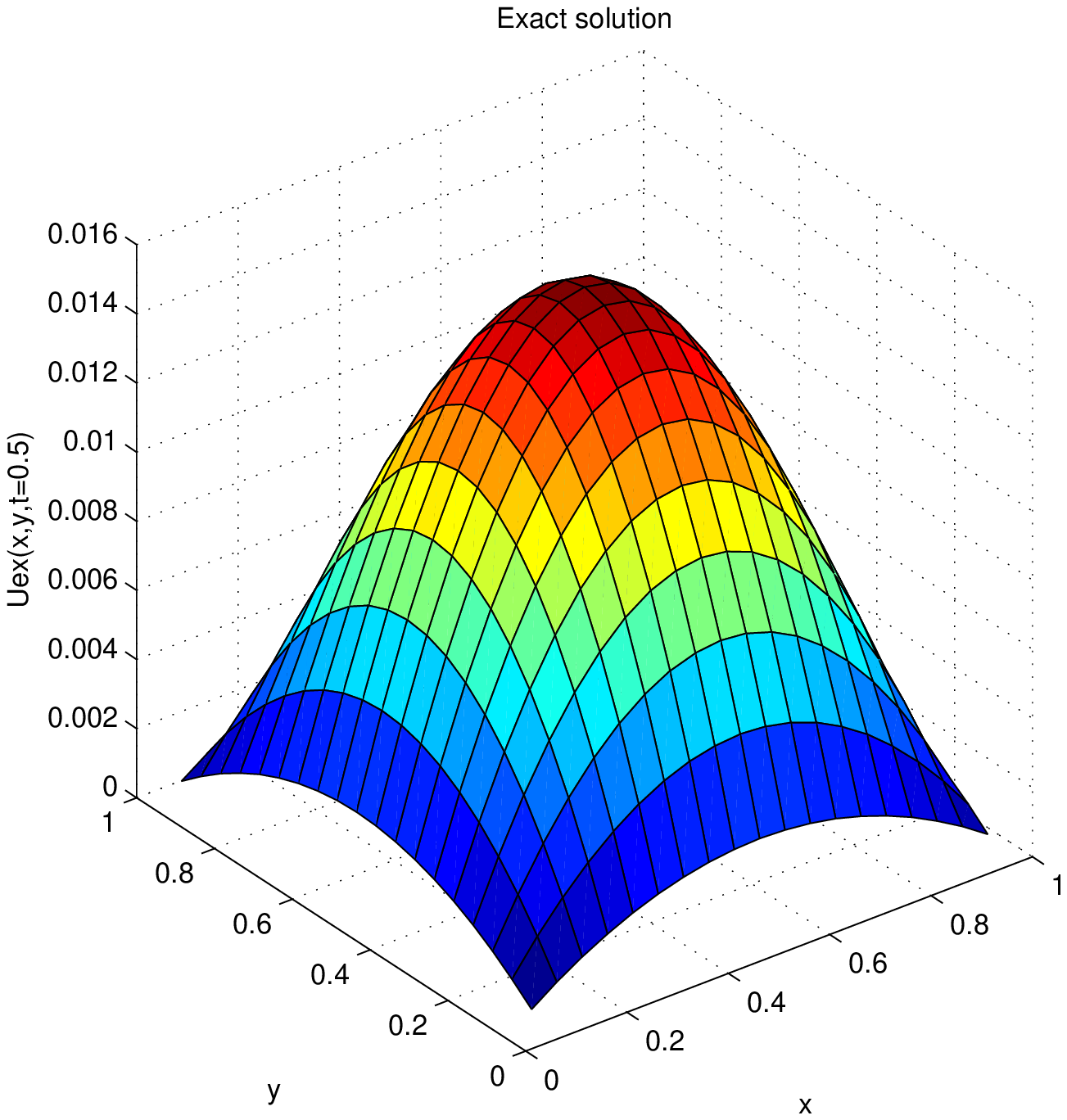}\label{exact}} }
\mbox{\subfigure {\epsfxsize 60mm\epsffile{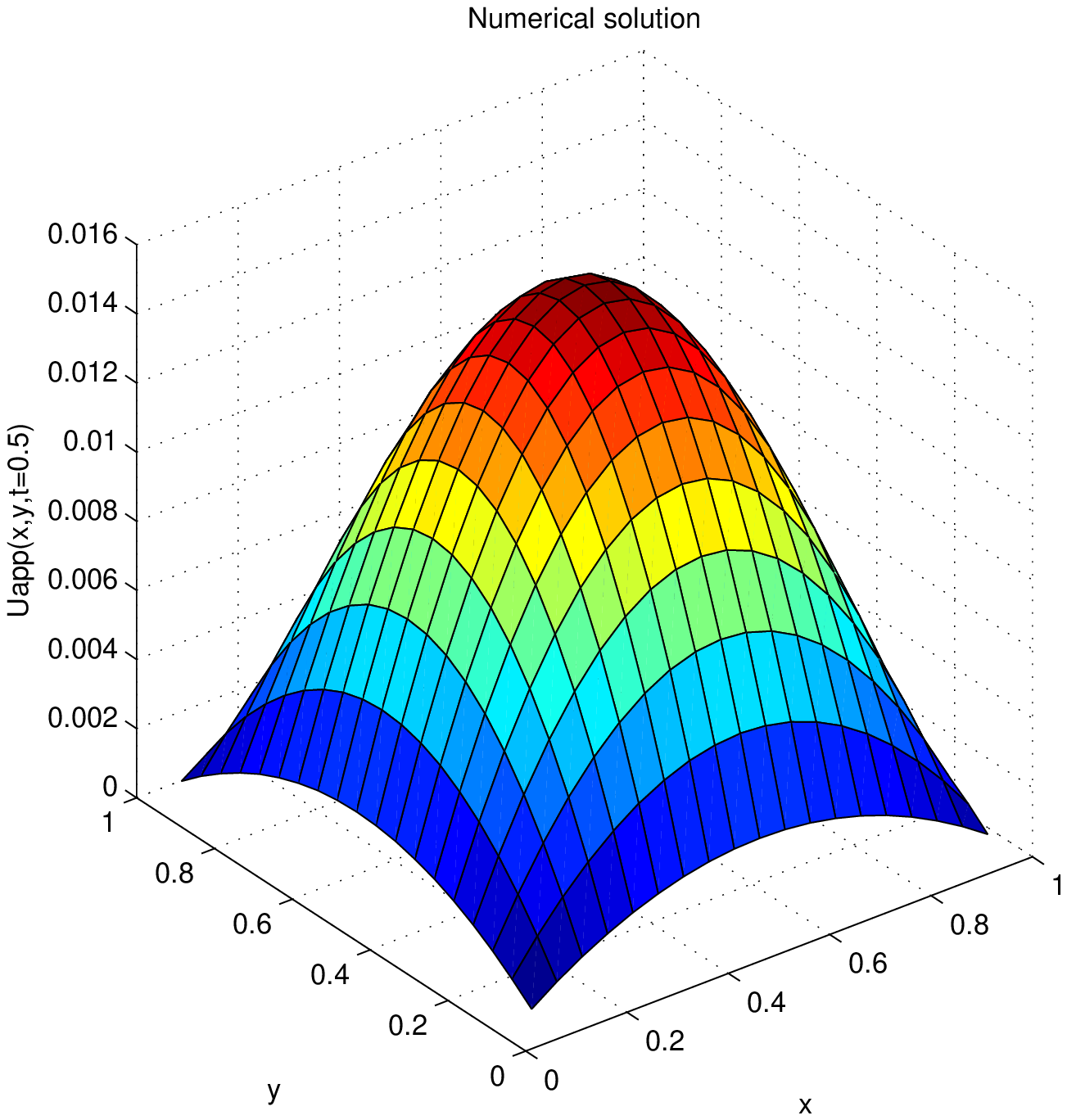}\label{1}}}
 \caption{Comparison of numerical and exact solution when $\alpha=0.9$} \label{comp4}
\end{figure}
From Figures \ref{comp1}-\ref{comp4}, we can see that the numerical results are in good accordance with exact solutions.

\textbf{Example 2} Consider the space fractional diffusion equation
\begin{equation}
\begin{aligned}
\frac{\partial u(x,y,t)}{\partial t}&=0.01\bigg(\frac{\partial^\beta u}{\partial |x|^\beta}+\frac{\partial^\beta u}{\partial |y|^\beta}\bigg)+f(x,y,t),\\
&\qquad 1<\beta<2,\quad 0\leq x\leq 1,~0\leq y\leq 1,~t\geq 0,
\end{aligned}
\end{equation}
together with initial and boundary conditions
\begin{equation}
u(x,y,0)=0,\quad 0<x<1,~0<y<1,
\end{equation}
\begin{equation}
\begin{aligned}
&u(0,y,t)=u(1,y,t)=0,\\
&u(x,0,t)=u(x,1,t)=0,
\end{aligned}
\end{equation}
where
\begin{equation*}
\begin{aligned}
f(x,y,t)&=2t(x-x^2)(y-y^2)+\frac{0.02t^2}{2\cos(\beta\pi/2)}\bigg\{(y-y^2)\\
&\qquad\bigg[\frac{x^{1-\beta}+(1-x)^{1-\beta}}{\Gamma(2-\beta)}\\
&\qquad\qquad-\frac{2x^{2-\beta}+2(1-x)^{2-\beta}}{\Gamma(3-\beta)}\bigg]\\
&\qquad+(x-x^2)\bigg[\frac{y^{1-\beta}+(1-y)^{1-\beta}}{\Gamma(2-\beta)}\\
&\qquad\qquad\qquad -\frac{2y^{2-\beta}+2(1-y)^{2-\beta}}{\Gamma(3-\beta)}\bigg]\bigg\}.
\end{aligned}
\end{equation*}
The exact solution of above equation is
$$u(x,y,t)=t^2 x(1-x)y(1-y).$$
 Letting $h_x=h_y=\frac{1}{20}$ as spatial step size, $\tau=\frac{1}{250}$ as time step length, the comparisons between numerical and exact solutions at time $t=1$ with different orders $\beta$ are shown as follows
\begin{figure}[!htb]
\centering \mbox{\subfigure{\epsfxsize 60mm\epsffile{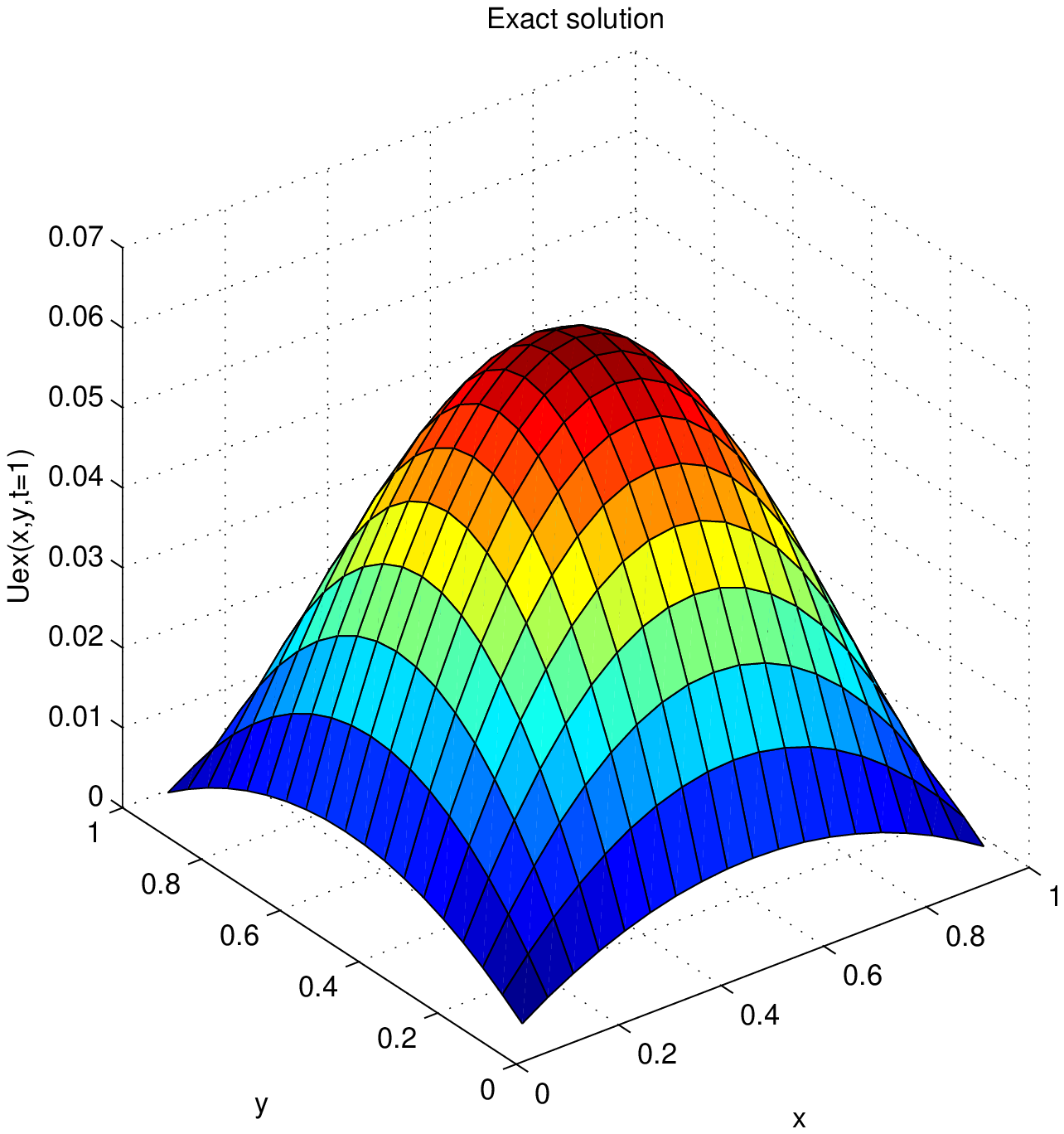}\label{exact}} }
\mbox{\subfigure {\epsfxsize 60mm\epsffile{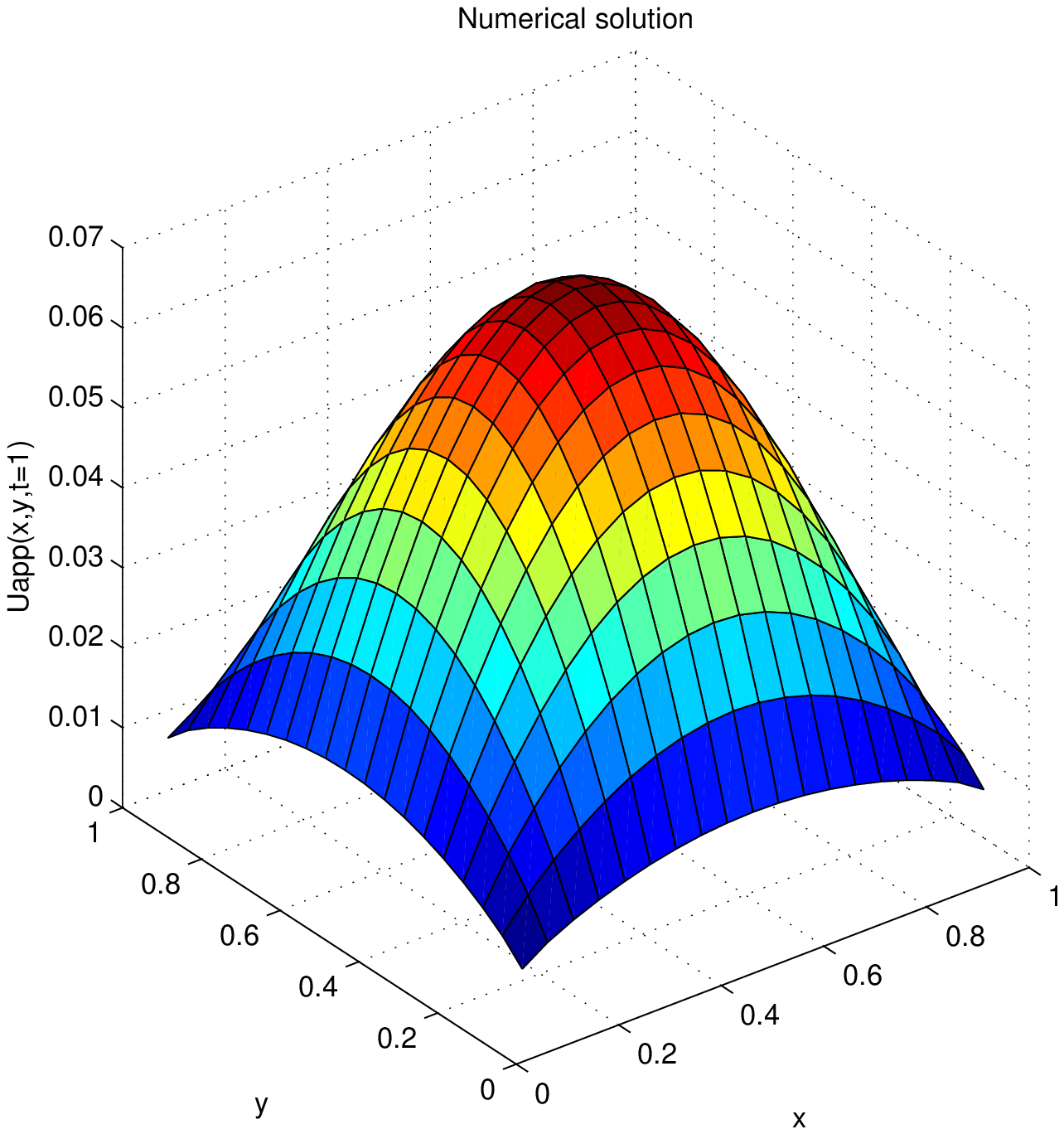}\label{1}}}
 \caption{Comparison of numerical and exact solution when $\beta=1.3$} \label{comp5}
\end{figure}

\begin{figure}[!htb]
\centering \mbox{\subfigure{\epsfxsize 60mm\epsffile{uexact.eps}\label{exact}} }
\mbox{\subfigure {\epsfxsize 60mm\epsffile{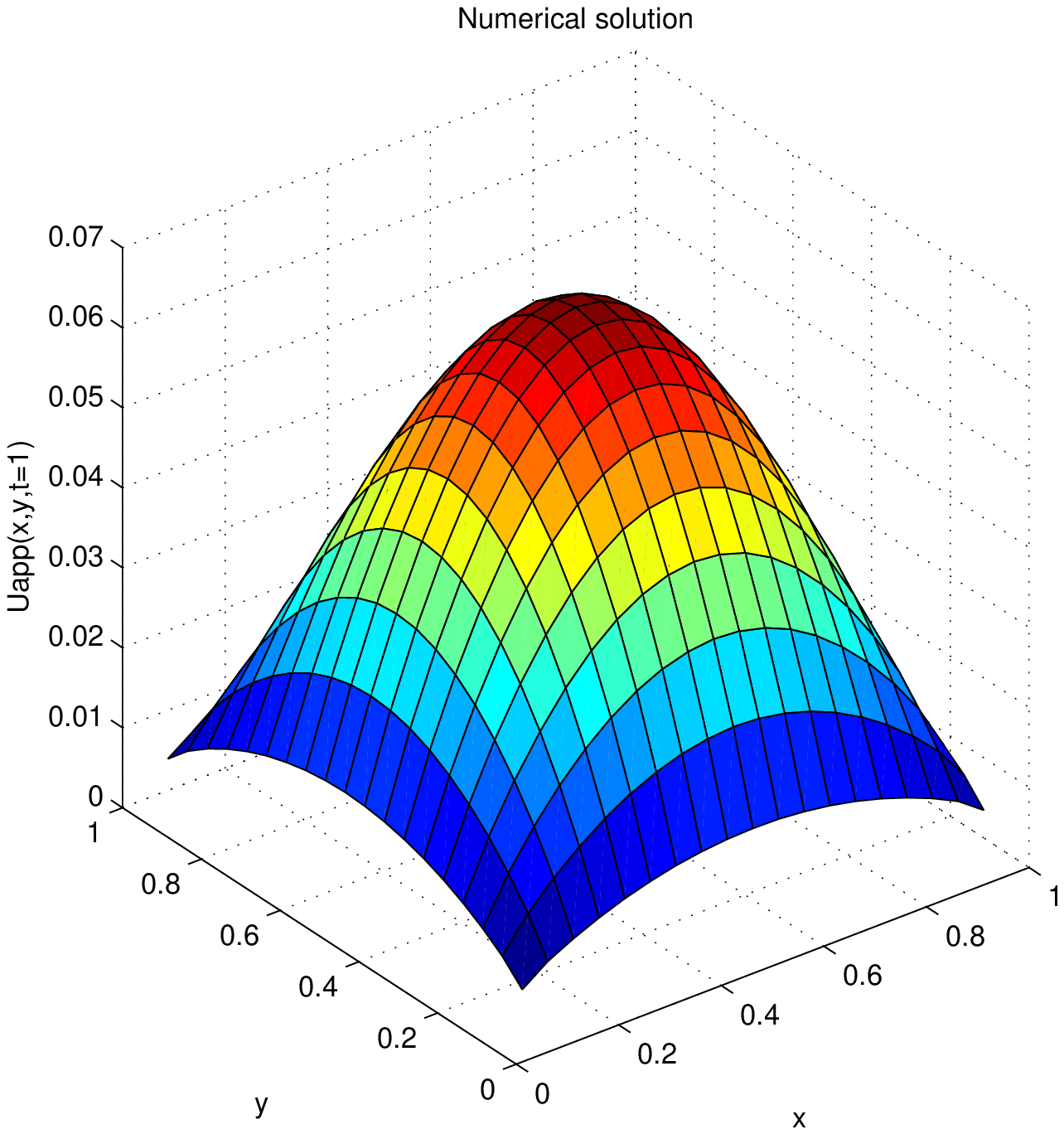}\label{1}}}
 \caption{Comparison of numerical and exact solution when $\beta=1.5$} \label{comp6}
\end{figure}

\begin{figure}[!htb]
\centering \mbox{\subfigure{\epsfxsize 60mm\epsffile{uexact.eps}\label{exact}} }
\mbox{\subfigure {\epsfxsize 60mm\epsffile{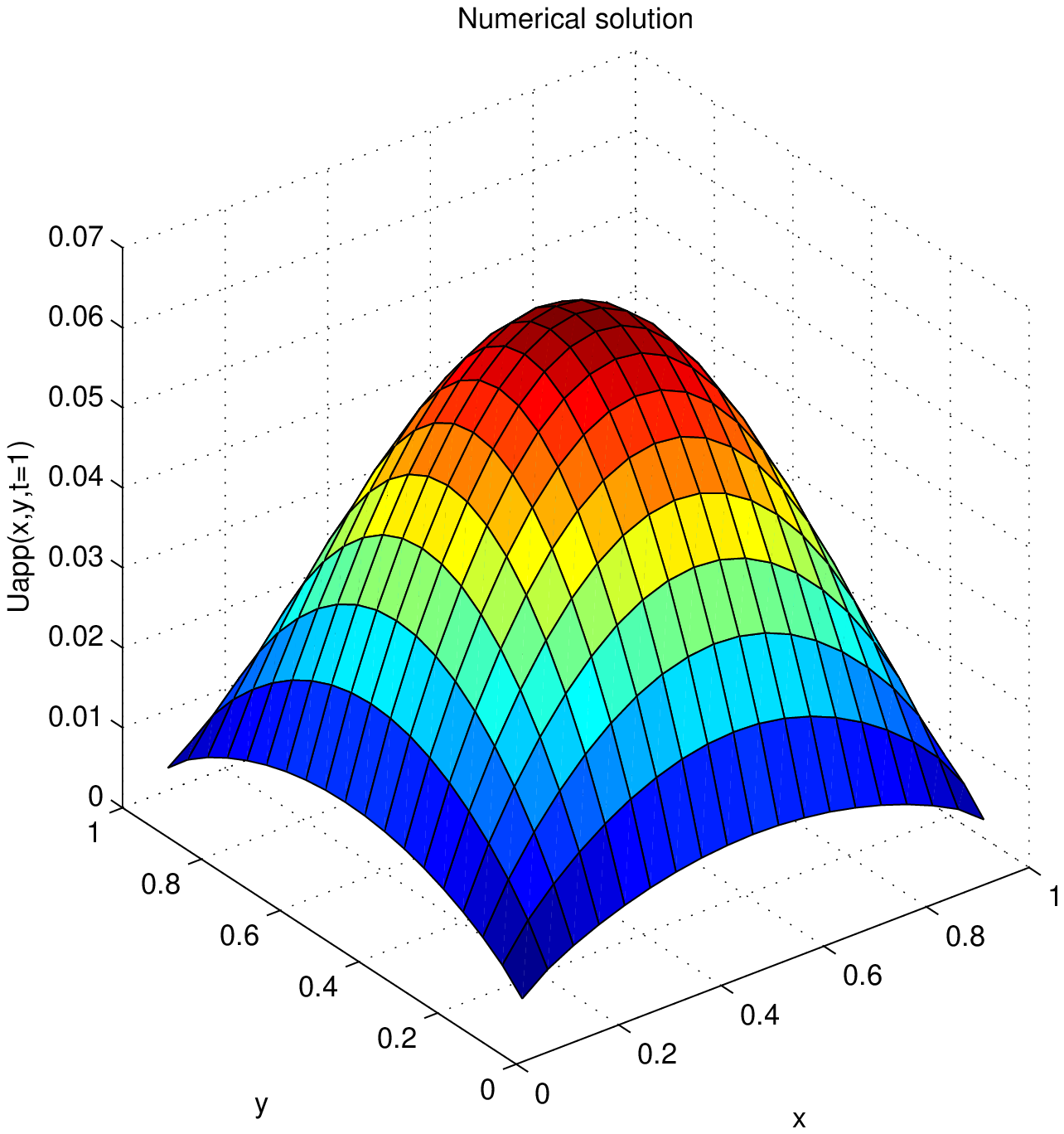}\label{1}}}
 \caption{Comparison of numerical and exact solution when $\beta=1.7$} \label{comp7}
\end{figure}

\begin{figure}[!htb]
\centering \mbox{\subfigure{\epsfxsize 60mm\epsffile{uexact.eps}\label{exact}} }
\mbox{\subfigure {\epsfxsize 60mm\epsffile{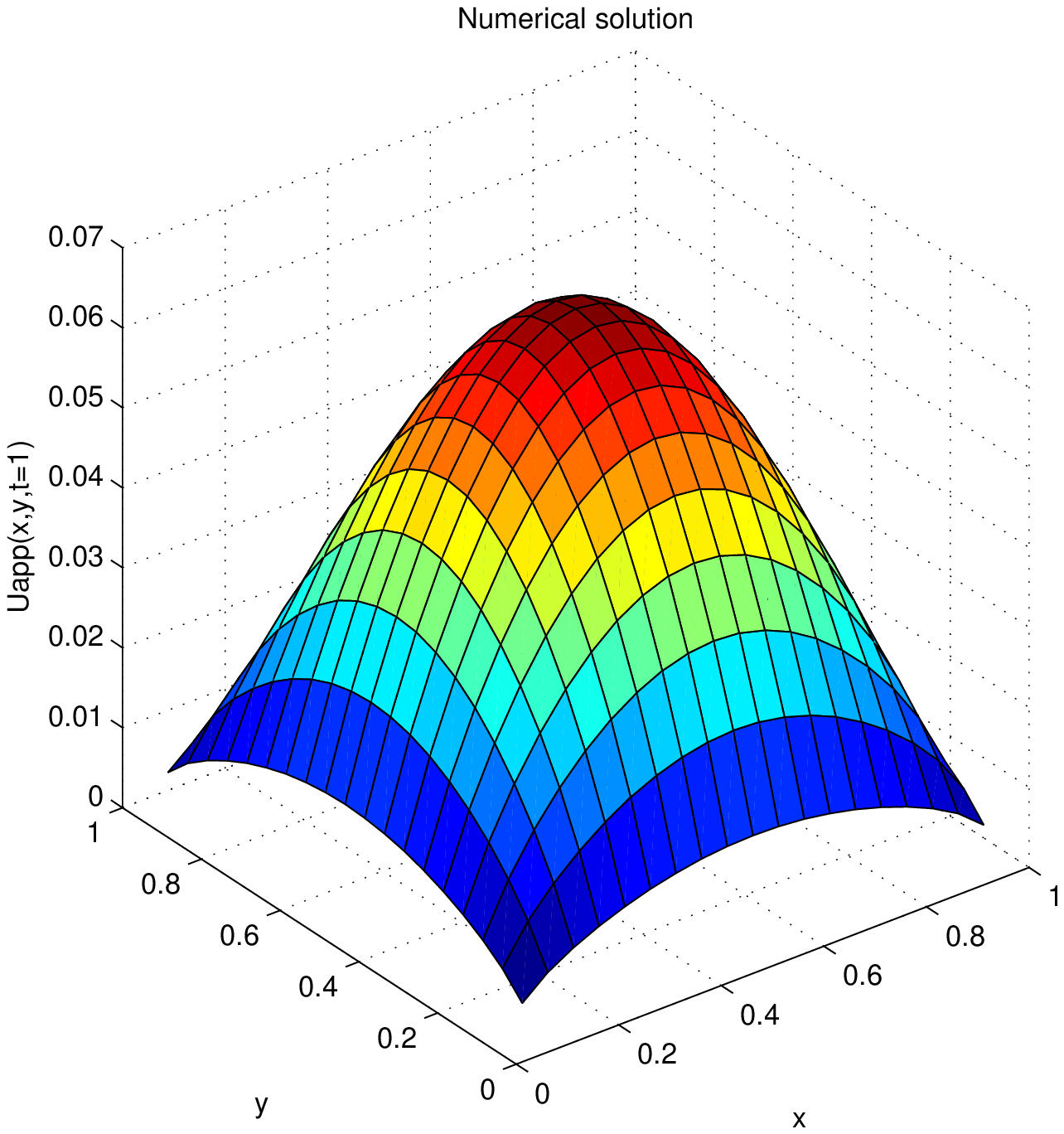}\label{1}}}
 \caption{Comparison of numerical and exact solution when $\beta=1.9$} \label{comp8}
\end{figure}
We can see that the numerical solutions are agree with the exact solutions.

From the above two examples, it is easy to see that  \texttt{FO-DiffMAS-2D} is efficient in solving two dimensional diffusion equations.
\section*{References}
\bibliographystyle{elsarticle-num}
\bibliography{acc15}

\begin{thebibliography}{10}
\expandafter\ifx\csname url\endcsname\relax
  \def\url#1{\texttt{#1}}\fi
\expandafter\ifx\csname urlprefix\endcsname\relax\def\urlprefix{URL }\fi
\expandafter\ifx\csname href\endcsname\relax
  \def\href#1#2{#2} \def\path#1{#1}\fi

\bibitem{metzler2000random}
R.~Metzler, J.~Klafter, The random walk's guide to anomalous diffusion: a
  fractional dynamics approach, Physics reports 339~(1) (2000) 1--77.

\bibitem{zarzhitsky2005swarms}
D.~Zarzhitsky, D.~F. Spears, W.~M. Spears, Swarms for chemical plume tracing,
  in: Swarm Intelligence Symposium, 2005. SIS 2005. Proceedings 2005 IEEE,
  IEEE, 2005, pp. 249--256.

\bibitem{demetriou2006power}
M.~A. Demetriou, Power management of sensor networks for detection of a moving
  source in 2-d spatial domains, in: American Control Conference, 2006, IEEE,
  2006, pp. 6--pp.

\bibitem{han2014multiple}
J.~Han, Y.~Chen, Multiple uav formations for cooperative source seeking and
  contour mapping of a radiative signal field, Journal of Intelligent \&
  Robotic Systems 74~(1-2) (2014) 323--332.

\bibitem{du1999centroidal}
Q.~Du, V.~Faber, M.~Gunzburger, Centroidal voronoi tessellations: applications
  and algorithms, SIAM review 41~(4) (1999) 637--676.

\bibitem{chen2007optimal}
Y.~Chen, Z.~Wang, J.~Liang, Optimal dynamic actuator location in distributed
  feedback control of a diffusion process, International Journal of Sensor
  Networks 2~(3) (2007) 169--178.

\bibitem{stark2013optimal}
B.~Stark, S.~Rider, Y.~Chen, Optimal pest management by networked unmanned
  cropdusters in precision agriculture: A cyber-physical system approach, in:
  Research, Education and Development of Unmanned Aerial Systems, Vol.~2, 2013,
  pp. 296--302.

\bibitem{liang2004diff}
J.~Liang, Y.~Chen, Diff-mas2d (version 0.9) user¡¯s manual: A simulation
  platform for controlling distributed parameter systems (diffusion) with
  networked movable actuators and sensors (mas) in 2d domain, CSOIS, Utah State
  University, Tech. Rep. USU-CSOIS-TR-04-03.

\bibitem{podlubny1998fractional}
I.~Podlubny, Fractional differential equations: an introduction to fractional
  derivatives, fractional differential equations, to methods of their solution
  and some of their applications, Vol. 198, Academic press, 1998.

\bibitem{ju2002probabilistic}
L.~Ju, Q.~Du, M.~Gunzburger, Probabilistic methods for centroidal voronoi
  tessellations and their parallel implementations, Parallel Computing 28~(10)
  (2002) 1477--1500.

\bibitem{chen2006optimal}
Y.~Q. Chen, Z.~Wang, K.~L. Moore, Optimal spraying control of a diffusion
  process using mobile actuator networks with fractional potential field based
  dynamic obstacle avoidance, in: Networking, Sensing and Control, 2006.
  ICNSC'06. Proceedings of the 2006 IEEE International Conference on, IEEE,
  2006, pp. 107--112.

\bibitem{chen2005actuation}
Y.~Chen, Z.~Wang, J.~Liang, Actuation scheduling in mobile actuator networks
  for spatial-temporal feedback control of a diffusion process with dynamic
  obstacle avoidance, in: Mechatronics and Automation, 2005 IEEE International
  Conference, Vol.~2, IEEE, 2005, pp. 752--757.

\bibitem{chen2005optimal}
Y.~Q. Chen, Z.~Wang, J.~Liang, Y.~Chen, Optimal dynamic actuator location in
  distributed feedback control of a diffusion process, in: Decision and
  Control, 2005 and 2005 European Control Conference. CDC-ECC'05. 44th IEEE
  Conference on, IEEE, 2005, pp. 5662--5667.

\bibitem{cortes2002coverage}
J.~Cortes, S.~Martinez, T.~Karatas, F.~Bullo, Coverage control for mobile
  sensing networks, in: Robotics and Automation, 2002. Proceedings. ICRA'02.
  IEEE International Conference on, Vol.~2, IEEE, 2002, pp. 1327--1332.

\bibitem{howard2002mobile}
A.~Howard, M.~J. Matari{\'c}, G.~S. Sukhatme, Mobile sensor network deployment
  using potential fields: A distributed, scalable solution to the area coverage
  problem, in: Distributed autonomous robotic systems 5, Springer, 2002, pp.
  299--308.

\bibitem{ortigueira2008fractional}
M.~D. Ortigueira, Fractional central differences and derivatives, Journal of
  Vibration and Control 14~(9-10) (2008) 1255--1266.

\end{thebibliography}
\end{document}